\documentclass[referee]{raa}           
\usepackage{graphicx,times}
\usepackage{natbib}
\usepackage{amssymb,amsmath}
\bibpunct{(}{)}{;}{a}{}{,}

\usepackage[a4paper=true,pagebackref=true]{hyperref}
\hypersetup{pdftitle = The title of my PDF, pdfauthor = My name, pdfsubject= The subject, pdfkeywords = keyword1 keyword2 keyword3} 
\hypersetup{colorlinks = true, linkcolor = green, anchorcolor = red, citecolor = blue, filecolor = red, pagecolor = red, urlcolor = red}

\begin{document}

   \title{The Ellerman bomb and Ultraviolet burst triggered successively by an emerging magnetic flux rope
$^*$
\footnotetext{\small $*$ }
}

 \volnopage{ {\bf 20XX} Vol.\ {\bf X} No. {\bf XX}, 000--000}
   \setcounter{page}{1}

   \author{Guanchong Cheng\inst{1,2,3}, Lei Ni\inst{1,3,4}, Yajie Chen\inst{5}, Udo Ziegler
      \inst{6},  Jun Lin\inst{1,2,4}
   }

   \institute{Yunnan Observatories, Chinese Academy of Sciences,Kunming, Yunnan 650216, P. R. China; {\it leini@ynao.ac.cn}\\
        \and
             University of Chinese Academy of Sciences, Beijing 100049, P. R. China.\\
	\and
%	  Center for Astrophysics, University of Science and Technology of China, Hefei 230026, China\\
CAS Key Laboratory of Solar Activity, National Astronomical Observatories, Beijing 100101, China\\
\and 
Center for Astronomical Mega-Science, Chinese Academy of Sciences, 20A Datun Road, Chaoyang District, Beijing 100012, P. R. China\\
\and
School of Earth and Space Sciences,Peking University, Beijing 100871, China\\
\and
Astrophysikalisches Institut Potsdam D-14482 Potsdam, Germany\\
\vs \no
   {\small Received 2021 4 25; accepted 2021 XX XX}
}

\abstract{Ellerman bombs (EBs) and Ultraviolet (UV) bursts are common brightening phenomena which are usually generated in the low solar atmosphere of emerging flux regions. In this paper, we have investigated the emergence of an initial un-twisted magnetic flux rope based on three-dimensional (3D) magneto-hydrodynamic (MHD) simulations. The EB-like and UV burst-like activities successively appear in the U-shaped part of the undulating magnetic fields triggered by Parker Instability. The EB-like activity starts to appear earlier and lasts for about 80 seconds. Six minutes later, a much hotter UV burst-like event starts to appear and lasts for about 60 seconds. Along the direction vertical to the solar surface, both the EB and UV burst start in the low chromosphere, but the UV burst extends to a higher altitude in the up chromosphere. The regions with apparent temperature increase in the EB and UV burst are both located inside the small twisted flux ropes generated in magnetic reconnection processes, which are consistent with the previous 2D simulations that most hot regions are usually located inside the magnetic islands. However, the twisted flux rope corresponding to the EB is only strongly heated after it floats up to an altitude much higher than the reconnection site during that period. Our analyses show that the EB is heated by the shocks driven by the strong horizontal flows at two sides of the U-shaped magnetic fields. The twisted flux rope corresponding to the UV burst is heated by the driven magnetic reconnection process.
}

   \authorrunning{Guanchong Cheng et al. }            %author_head in even pages
   \titlerunning{EBs and UV bursts}  % title_head in odd pages
   \maketitle

%________________________________________________ sections below
% 
\section{Introduction}           %% first-level sections will be auto-capitalized
\label{sect:intro}

The emergence of the magnetic flux is one of the most important dynamic processes around the solar surface. The strength of magnetic fields emerging from solar interior could reach several hundreds to thousands of G in the solar surface\citep[e.g.,][]{Getling2019,Liu2020,Leenaarts2018,Yan2020,Yan2017}. Different kinds of transient brightenings observed in the low solar atmosphere usually connect with the flux emergence process \citep[e.g.,][]{Xue2016,Zhao2017,Huang2018,Tian2018a,Huang2019,Yang2019}. EBs and UV bursts are two kinds of the most common ones, which are usually considered to form in a magnetic reconnection process triggered by the emerging of a magnetic flux tube or when a flux tube approaches the background magnetic fields with an opposite direction \citep{Pariat2004,Hashimoto2010}. The obvious magnetic cancellation in the photosphere is usually observed when these reconnection events happen \citep[e.g.,][]{Wang1995,Pariat2004,Pariat2007,Hashimoto2010,Peter2014,Zhao2017,Tian2018b}.

EBs are originally named as ``solar hydrogen bombs'' by \cite{Ellerman1917}, The shape of their H$\alpha$ spectral line profile looks like a moustache. Usually, the maximum emission is at around H$\alpha$ $\pm$ 1 \AA\ and the emission is gradually disappearing at $\pm$ 5\AA\ \citep{Severny1968}. The observations show that EBs also have strong emissions in Ca II H and IR lines, while there is no obvious signature in Na I D1 and Mg I b2 lines \citep[e.g.,][]{Rutten2015,Vissers2013,Vissers2015}. The typical size of EBs is about $1^{\prime\prime}$ and their typical life time is from a few minutes to tens of minutes \citep[e.g.,][]{Georgoulis2002}. The released energy in an EB is usually about $10^{26}$-$10^{28}$ erg \citep{Fang2006}. On the basis of the semi-empirical model of numerous combinations of the spectral lines, the estimated temperature increase in EBs is about a few $10^{2}$ to $10^{3}$ K \citep[e.g.,][]{Nelson2015,Grubecka2016,Hong2017a,Hong2017b}.

UV bursts are another kind of transient events which are formed in the low solar atmosphere and first discovered by the Interface Region Imaging Spectrograph satellite \citep[IRIS;][]{De2014}. They have strong emissions in Si IV lines and also have responses in the UV continua at 1600 \AA\ and 1700 \AA\ \citep[e.g.,][]{Peter2014, Tian2016, Chitta2017, Guglielmino2019, Chen2019}. Strong absorbtion in Ni II lines is usually observed, which indicates that part of the UV burst is located at fairly low altitudes below the middle chromosphere. In the dense photosphere environment, the generation of Si IV emission needs a temperature increase of about 20,000 K. If the UV bursts are formed in the upper chromosphere where the plasma density is much lower than the photosphere, a temperature increase of about 80,000 K is needed to form the Si IV emission \citep{Rutten2016}. The spectral line profile of Si IV in UV burst is significantly enhanced and broadened\citep{Peter2014,Tian2016}, the red and blue wings of the line profile of Si IV  emission lines can reach above $100$\,km s$^{-1}$  away from line center. However, \cite{Hou2016} found that some UV bursts have narrow line widths which are smaller than 20 km\,s$^{-1}$. The joint observations of ground-based telescopes and the IRIS satellite have discovered that UV bursts are sometimes coexistent with EBs \citep[e.g.,][]{Tian2016}. When the two phenomena exist together, they occur almost simultaneously or the UV emissions appear several minutes later than the EBs.\citep{Ortiz2020}. \cite{Chen2019} have identified 161 EBs based on the high resolution observations from the Goode Solar Telescope(GST), they found that 20 of them have observed features of UV bursts according to the associated IRIS observations. Most UV bursts associated with EBs tend to occur in the upper part of the EBs\citep{Chen2019}.  

Some typical features of EBs have been well simulated by using the single-fluid MHD model \citep[e.g.,][]{ChenPF2001,Isobe2007,Archontis2009,Xu2011, Danilovic2017a}. \cite{ChenPF2001} tested several cases with different plasma densities, their results indicated that EBs can be formed in a magnetic reconnection process at different heights from photosphere to the middle chromosphere. The Parker instability has been applied to triggering flux emergence, then the multiple emerging loops expand and interact with one another, magnetic reconnection happens in the U-shaped like magnetic field structures \citep{Isobe2007,Archontis2009}, such a scenario agrees well with the serpentine field lines observed in bald patches where EBs appear \citep[e.g.,]{Pariat2004}. Though the temperature increases of hundreds of to several thousands K in those simulations \citep[e.g.,][]{Archontis2009} were just perfect to reproduce the brightenings at both H$\alpha$ wings, further quantitative analysis comparing with observations were not completed because of the lack of radiative cooling process. Recently, \cite{Danilovic2017a} studied the EB-like events by using the 3D Radiation Magnetohydrodynamics (RMHD) code that includes radiative transfer and radiative losses, the authors compared their results with corresponding high-resolution observations and showed that the emerging serpentine-like magnetic filed lines indeed lead to the formations of the EB-like events for the first time. Their further simulations presented the cases both in the quiet Sun and active region, and showed their similarities \citep{Danilovic2017b}. The results indicated that the flame-like morphology is related to the intricacies of the ongoing reconnection as well as the orientations of sight lines, and they concluded that the EB features are caused by reconnection of strong-flied patches of opposite polarity in the regions where the surface flows are the strongest \citep{Danilovic2017b}. Their analyses \citep{Danilovic2017a, Danilovic2017b} were limited to the temperature minimum region (TMR) and below because their radiative cooling model does not include  the strongest chromospheric lines and non-equilibrium ionization effects. 

For the first time, \cite{Ni2015} showed that the plasmas near the solar TMR can be heated above tens of thousand K when the reconnecting magnetic fields reaches above several hundred G. UV bursts could only be  formed during a magnetic reconnection process with a small plasma $\beta$ ($<$1) \citep[e.g.,][]{Ni2016, Ni2018a, Peter2019}. The RMHD code Bifrost has been used to model UV bursts \citep[e.g.,][]{Hansteen2017, Hansteen2019}, the optically thick radiative transfer and radiative losses are included from the photosphere to the low chromosphere and the 1D non-local thermodynamic equilibrium (non-LTE) radiative table \citep{Carlsson2012} is applied above the low chromosphere in these simulations. The synthesized Si IV and Mg II spectrum lines are consistent with the observational ones \citep[e.g.,][]{Rouppe2017,Hansteen2019}. The UV emissions in those simulations are basically above the middle chromosphere. However, the recent 2.5D high resolution MHD simulations \citep{Ni2021} with a more realistic magnetic diffusion and partially ionized effect showed that the UV burst can be formed during a reconnection process with stronger magnetic fields ($\sim$ 500 G) in the low chromosphere, where the plasma density is about two orders of magnitude higher than those previous simulations \citep[e.g.,][]{Hansteen2017}. \cite{Hansteen2019} has numerically studied the coexisting EBs and UV bursts, they conclude that the lower cool part and the hot upper part of a vertical long current sheet correspond to the EB and UV burst in these coexisting events, respectively. The results in \citep{Ni2021} showed that the EB indeed extends to a lower altitude in the photosphere in such a coexisting event, but  the hot UV emissions and parts of the much cooler plasmas can be concentrated in one plasmoid or the turbulent reconnection region at about the same altitude in the low chromosphere. 

Since magnetic reconnection is considered as the main mechanism to trigger the formations of UV bursts and EBs, the further 3D high resolution simulations by including more realistic diffusivities and partially ionization effects are very important to reveal the magnetic reconnection and heating mechanisms in these low atmosphere activities. Then, we can better understand the formations of UV bursts and EBs and their relationships and diminish the differences between the observations and simulations. The previous high resolution 2D simulations showed that the hot plasmas in the reconnection region are usually located inside the plasmoids (also named as magnetic islands) \citep[e.g.,][]{Ni2015, Ni2016}, the plasmoid instability is also one of the main mechanisms to lead fast magnetic reconnection in the low solar atmosphere \citep[e.g.,][]{Ni2015, Ni2018b}. The non-LTE inversions of the low-atmosphere reconnection based on SST and IRIS observations indicated the existence of the plasmoids \citep{Vissers2019}. However, non of the previous 3D simulations of UV bursts or EBs clearly showed the newly generated flux ropes (corresponding to magnetic islands in 2D) and focused on the fine structures in the magnetic reconnection region.

In this paper, we investigate the emergence of a single flux rope based on the 3D MHD simulation calculated by the NIRVANA3.8 code \citep{Ziegler2008,Ziegler2011}. A special initial perturbation of density triggers the emerging process of the magnetic flux tube. The EB and UV burst are found to be formed successively in the U-shaped part of the undulating magnetic fields, and we analyze the fine structures in the magnetic reconnection sites. The radiative transfer code Multi 1.5 D has been used to synthesize the H$\alpha$ images and spectral line profile of the EB \citep{Carlsson1986}. We also synthesize the Si IV emission and spectral line profile of the UV burst based on our numerical results. The heating mechanisms of these two events are also analyzed and revealed, and we find that the EB in this simulation is not directed heated by magnetic reconnection.  Section 2 shows the numerical models and methods. The numerical results are presented in Section 3. Section 4 describes the summaries and discussions.

\section{Models and Methods} \label{sec:style}
\subsection{MHD equations} \label{subsec:tables}
The solved MHD equations in the this work are as follows:

\begin{eqnarray}
\frac{\partial \rho }{\partial t}&=&-\nabla \cdot \left (\rho \mathbf{v}\right ) \\
\frac{\partial \left ( \rho \mathbf{v} \right )}{\partial t} &=&- \nabla \cdot \left [ \rho \mathbf{vv} +\left ( p+\frac{1}{2\mu _{\rm 0}}\left | \mathbf{B} \right |^{2} \right )I-\frac{1}{\mu _{\rm 0}}\mathbf{BB} \right ] \nonumber \\
&& +\rho \mathbf{g} \\
\frac{\partial e}{\partial t}&=&-\nabla \cdot\left [ \left ( e+p+\frac{1}{2\mu _{\rm 0}}\left | \mathbf{B} \right |^{2} \right )\mathbf{v} \right ]  \nonumber \\
&&+\nabla \cdot\left [ \frac{1}{\mu _{\rm 0}}\left ( \mathbf{v}\cdot \mathbf{B} \right )\mathbf{B}   \right ] \nonumber \\
%&&+\nabla \cdot \left [ \frac{\eta }{\mu _{0}}\mathbf{B}\times \left ( \nabla \times \mathbf{B} \right )\right ]       \nonumber \\
&&+\rho \mathbf{g}\cdot \mathbf{v}+L_{\rm rad}+H  \label{enereq} \\
\frac{\partial \mathbf{B}}{\partial t} &=& \nabla \times \left ( \mathbf{v}\times \mathbf{B} \right )  \label{indeq} \\
e&=&\frac{p}{\gamma -1}+\frac{1}{2}\rho \left | \mathbf{v} \right |^{2}+\frac{1}{2\mu _{\rm 0}}\left | \mathbf{B} \right |^{2} \\
p&=&\frac{\left ( 1+Y_{\rm i} \right )\rho}{m_{\rm i}}k_{\rm B}T
\end{eqnarray}
where $\rho$, $\mathbf{v}$, $e$ and $\mathbf {B} $ are the plasma density, the fluid velocity, the total energy density and the magnetic field. $m_ {i} $, $p$, $T$ and $Y_{\rm i}$ represent the mass of an ion, the plasma pressure, the temperature, and the ionization degree of the plasmas. $\mathbf{g}=273.93$ m$\cdot$ s$^{-2}$ is the gravitational acceleration of the Sun. The vacuum permeability is set to $\mu_{\rm 0}$ = 4$\pi \times$ 10$^{-7}$ N$\cdot$ A$^{-2}$, $\mathbf{I}$ represents the unit tensor, $\gamma = \frac{5}{3}$ is the ratio of specific heats, $k_{\rm B}$ = 1.3806$\times 10^{-23}$ J$\cdot$ K$^{-1}$ is the Boltzmann constant.

Since the radiative transfer process is very important in the low atmosphere, the radiative cooling model proposed by \cite{Gan1990} is applied in our numerical simulation. The equation for this model is as follows:

\begin{eqnarray}
L_{\rm rad} &=& -1.547\times 10^{-42}Y_{\rm i}\left(\rho/m_{\rm i}\right)^{2}\alpha T^{1.5}\\
\alpha &=& 10^{c1} +2.37\times 10^{-4} e^{c2}  \nonumber \\
c1 &=& 2.75\times 10^{-6}y -5.45    \nonumber   \\
c2 &=& -y/163\times 10^{-3}
\end{eqnarray}

They derived their model on the basis of detailed non-LTE calculations. They found that the radiative loss calculated from their model is similar to that from the non-LTE calculation, using the distributions of plasma parameters from the solar atmosphere model (VALC) \citep{Vernazza1981}. Hence, it is reasonable to choose this model as an approximation for the radiative cooling in the lower solar atmosphere. The initial conditions should make the system to be in equilibrium at the beginning. Otherwise, some unphysical results will appear in the simulations. Therefore, we choose a heating function to make the energy equation to be in equilibrium at the beginning. Its formula is as follow:   

\begin{equation}
\begin{array}{l}
H = 1.547 \times 10^{-42} Y_{\rm i} \rho_{\rm 0} \rho \alpha {T_{\rm 0}}^{1.5}/m_{\rm i}^2
\end{array}
\end{equation}
where $\rho_{\rm 0}$ and the $T_{\rm 0}$ are the plasma density and temperature at the initial moment, respectively. Both the radiative cooling and heating functions are only included above the solar surface ($y > 0$\,km). We have tested a case by excluding the heating function, we find that the very cold plasmas ($T < 2000$\,K) appear in a much large area near the up boundary in this case. The case with the heating function makes the situation better. Though the cold plasmas still appear above the EB (as shown in section 3.1), but the temperature in the other regions near the up boundary is more realistic. What is more, excluding the heating term makes the system to be more unstable during the later stage, and the temperature of the UV burst region is even higher. However, the main conclusions in this work will not change by excluding the heating term.

Since the heat conduction effect is not efficient in the reconnection region in the solar photosphere and chromosphere\citep{Ni2021}, we dropped the heat conduction term in this work. Ambipolar diffusion might be very important above the solar TMR region. However, the limited resolution makes the numerical diffusion to be larger than the ambipolar diffusion and physical magnetic diffusion in the chromosphere. Therefore, we dropped the terms relating to the ambipolar diffusion and physical magnetic diffusion in this work, and numerical diffusion triggers magnetic reconnection just as some of the previous works \citep{Hansteen2017, Hansteen2019} . 

\subsection{Initial and boundary conditions} \label{subsec:tables}

\begin{figure*}
	\centering{\includegraphics[width=0.6\textwidth]{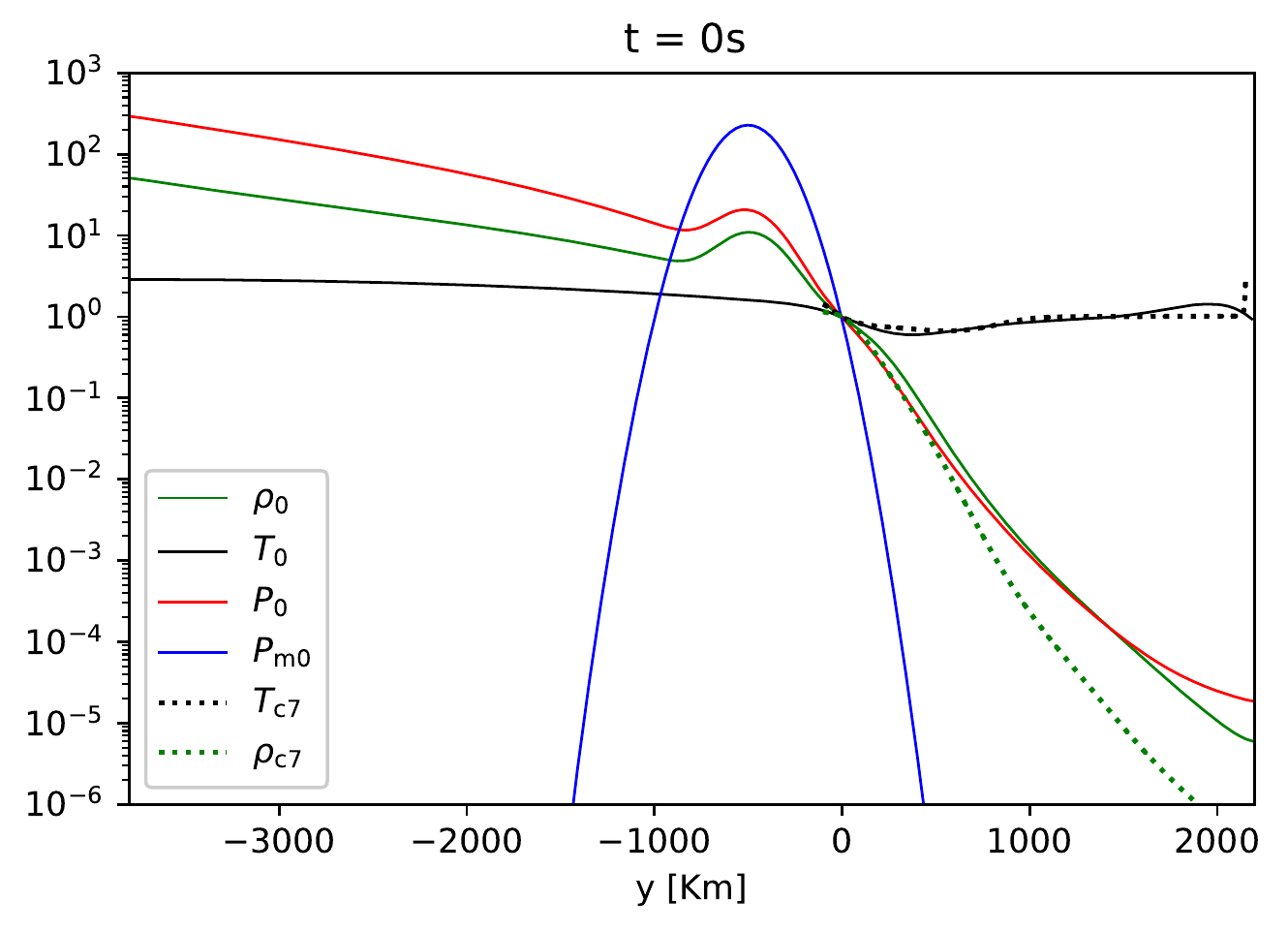}}
	\caption{ Distributions of temperature, density, thermal pressure, magnetic pressure along the y-direction. The variables are normalized by using the references values at y=0 km. These reference values are $\varrho_{\rm ref}=10^{-3.72}$ kg\,m$^{-3}$, $T_{\rm ref}=10^{3.84}$\,K, $P_{\rm ref}=10^{4.05}$\,Pa, $Pm_{\rm ref}=10^{3.6}$\,Pa. The dotted black line and the dotted green line represent the distributions of the temperature and plasma density in the C7 atmosphere model, respectively.}	
\end{figure*}

In this work, the simulation domain is from $-3.0$\,Mm to $3.0$\,Mm both in the $x$ and $z$ directions (horizontal directions), and it extends from $-3.8$\,Mm to $2.2$\,Mm in the $y$-direction that is perpendicular to the solar surface. The domain $-3.8$\,Mm$  < y < 0$\,Mm represents the convection zone of the Sun and $0$\,Mm $< y < 2.2$\,Mm represents the low solar atmosphere. There are 192 uniform grids in each direction, which means that the grid size is $31.25$\,km in our simulation. First of all, we ran a pure 2D hydrodynamic (HD) simulation by excluding the terms with magnetic fields. The initial velocities are set to be zero and the plasma parameters are uniform in the $x$-direction in this pure HD simulation. Since the initial magnetic field is zero, the initial system will not be in equilibrium if we use the exact plasma parameters as in the VALC model or the C7 model \cite{Avrett2008} (similar as the VALC model but have some improvements). On the basis of the C7 model, the initial atmosphere parameters above the solar surface ($y \ge -0.1L_0$, where $L_0=10^6$\,m) in our simulations are modified to satisfy the initial equilibrium. Firstly, we use the 4th degree polynomial to fit the temperature distribution in the C7 model to get the initial temperature distribution above the solar surface. We also use an analytical expression to fit the distribution of the ionization degree in the C7 model to get the initial ionization degree above the solar surface. It is hard to know the distributions of the plasma parameters below the solar surface ($y < -0.1L_0$), we choose a cos function to make the initial temperature distribution in this region to agree with the standard solar model, and we also simply fits the distribution of the ionization degree in this region by using a $\cos$ function. According to the initial distributions of temperature and ionization degree and the mass density at $y=-0.1L_0$ in the C7 model, we derived the initial mass density distribution in our simulation by solving the hydrostatic equilibrium equation $\nabla p=-\rho g$. The derived expressions for the distributions of these initial plasma parameters along the $y$-direction are as follow:

\begin{eqnarray}
T_{\rm 00}(y) &=& 
\begin{cases}
1.06\times10^4 \cos\left[0.424\left (y/L_0+3.80 \right) \right] + 9.38\times10^3 &
\textrm{$, y < -0.1L_0$}\\
\\
-1.23\times10^3 \left( y/L_0 \right)^{5}+9.76\times10^3 \left(y/L_0 \right)^{4}-2.84\times10^4 \left( y/L_0 \right)^{3} \\
+3.62\times10^4 \left( y/L_0 \right)^{2}-1.72\times10^4 \left( y/L_0 \right)+6.91\times10^3 &
\textrm{$, y \ge -0.1L_0$}
\end{cases}
\\
Y_{\rm i}(y) &=&
\begin{cases}
\cos\left[0.424\left(y/L_0+3.80\right)\right]  &
\textrm{$, y < -0.1L_0$}\\
\\
3\times 10^{-5} \frac{\exp \left[ 3.49 \left( y/L_0+0.12 \right)^{1.4}+
	2.40 \left(y/L_0+0.12 \right)^{0.9}+\left( y/L_0+0.12 \right)^{0.3} \right] } {17\left( y/L_0+0.12 \right)^{2.2}+2.64\left( y/L_0+0.12 \right)^{1.8}+0.006}  &
\textrm{$, y \ge -0.1L_0$}\\
\end{cases}
\\
\rho_{\rm 00}(y) &=& -7.98\times10^{-6}\left( y/L_0 \right)^{5}-1.69\times 10^{-5} \left( y/L_0 \right)^{4}+2.52\times 10^{-6} \left( y/L_0 \right)^{3} \nonumber \\
&  &+3.47\times 10^{-4}\left( y/L_0 \right)^{2}-5.25\times 10^{-6} \left( y/L_0 \right)+1.72\times 10^{-4}
\end{eqnarray}

From above expressions, one can find a mimicked TMR appears at around y=400 km, where the temperature is  $\sim 4300$ K. The range of the ionization is from $\sim10^{-4}$ to $\sim 1$, and the density decreased by about seven orders of magnitude from the bottom to the top of the simulation domain. After the HD simulation runs a physical time of $993$\,s,  the whole system tends to be in a smooth and stablest state. Then we use the plasma parameters at this time as our initial conditions to start our 3D MHD simulations. The distributions of these plasma parameters along $y$-direction are presented in Figure 1. The parameters are normalized by using the reference values of $\rho_{\rm ref}$, $T_{\rm ref}$, $P_{\rm ref}$, respectively. $\rho_{\rm ref}$, $T_{\rm ref}$, $P_{\rm ref}$ are the values at $y=0$\,km. The initial plasma parameters are uniform in both $x$ and $z$ directions. We should mention that the ionization degree is not time-dependent in our simulations.     
\begin{figure}
	\begin{center}
		\includegraphics[width=14.0cm]{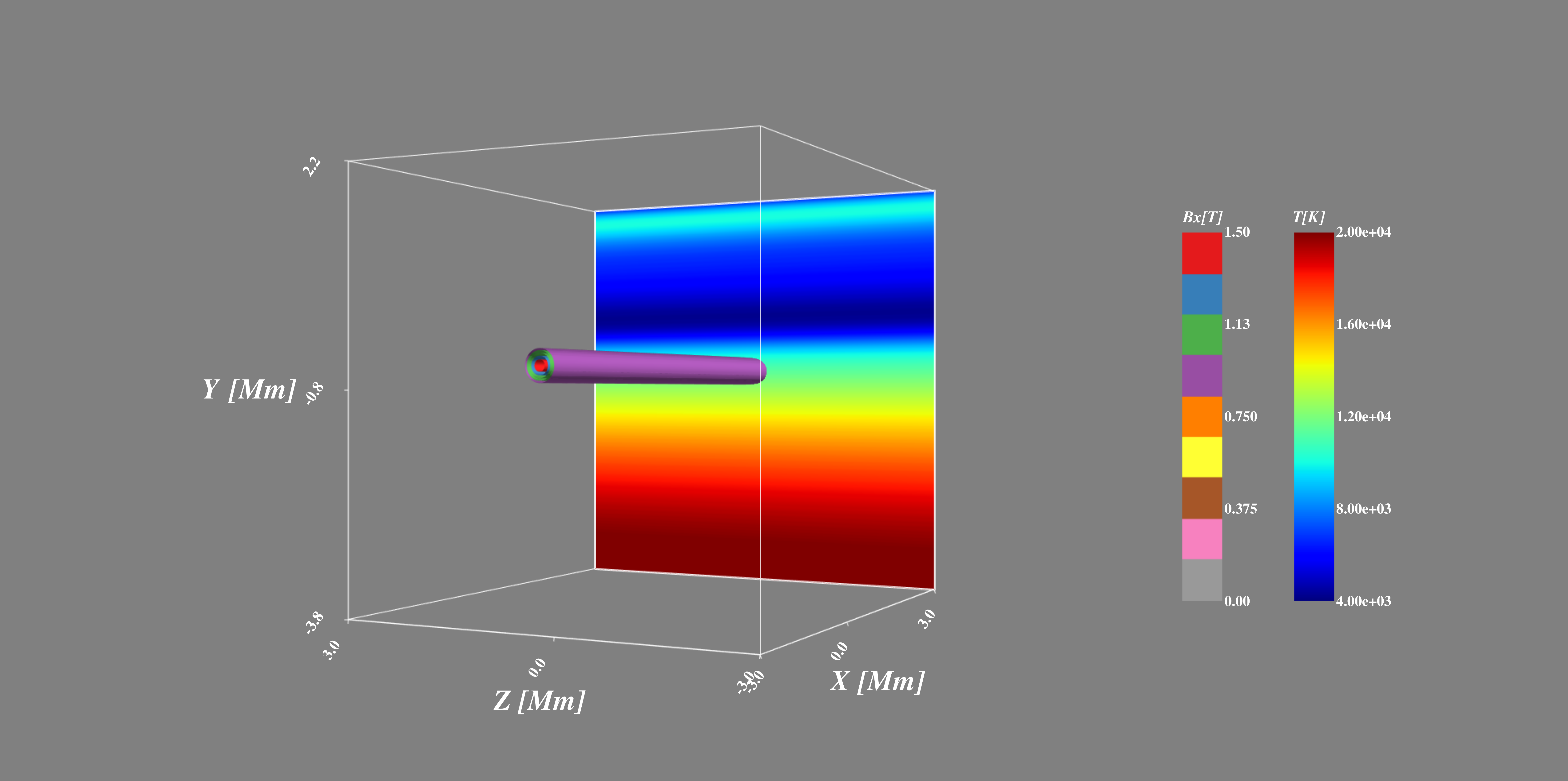}
	\end{center}
	\caption{ The three-dimensional overview map of the initial conditions, the slice is the temperature distribution in $y$-$z$ plane at $x=3000$\,km and the colorful tube represents the initial magnetic fields.}	
\end{figure}

At beginning of our MHD simulation, a small untwisted magnetic flux tube is embedded below the solar surface, its axis is along $x$-direction and located at $y$, $z$=[$-0.5$ Mm, $0$ Mm]. The magnetic field strength decays expontionally along the radial direction, $b_{\rm 0}=1.5$ \,T is the magnetic field strength at the center of flux tube. The formulas of the initial magnetic fields are as follow:
\begin{equation}
B_{\rm x0}  = b_{\rm 0}\exp\left[-\frac{ \left( y+5\times 10^{5} \right)^{2}+z^{2}}{ \left(3\times 10^{5} \right)^{2}}\right],
B_{\rm y0}=0,
B_{\rm z0}=0.
\end{equation}

The three-dimensional overview figure of the initial simulation domain is presented in Figure 2. The vertical slice is the initial temperature distribution in $y$-$z$ plane at $x=3000$\,km, the temperature is uniform along the $x$ and $z$ directions. The colorful tube represents the strength of the initial magnetic filed in the $x$-direction. There is no initial velocity driving. Such an initial flux tube can not rise high enough to trigger magnetic reconnection around the solar TMR if there are no additional perturbations in our simulations. The instabilities of magnetic buoyancy are essential for the emerging process of magnetic flux\citep{Matsumoto1993}. They are very important for driving the further rise of magnetic fields from the bottom of photosphere into the upper atmosphere. The arising of different modes depend on the angle between the magnetic filed B and the wavevector of perturbation k \citep{Kruskal1954,Newcomb1961}. When k // B, it is named as undular mode. In this mode the field lines undulate and the plasma slide down along the magnetic fields from the crests to the troughs, thus rising crests get lighter and the troughs get heavier, which facilitate the amplification of the perturbation \citep{Cheung2014}. The undular mode is also named as Parker Instability \citep{Parker1966}. 

Similar to the previous work \citep{Syntelis2015}, we add an initial density perturbation to trigger the Parker instability. The formula of the initial perturbation is as follow:

\begin{equation}
\Delta\rho = \frac{0.1}{2 \mu_{\rm 0}} \mathbf{B}^{2} p^{-1} \rho_{\rm 0} \left[ \arrowvert \cos \left( \frac{\pi x}{2L_0} \right) \arrowvert+ \exp\left(-\frac{x^{2}}{L_0^{2}}\right)\right].
\end{equation} 

This perturbation is coupled with the magnetic field, and it is basically located inside the flux tube. We find that the perturbation makes that the Parker instability to be satisfied \citep{Archontis2004} and the magnetic flux tube to rise from the convection zone to the lower atmosphere successfully. After this simulation starts, that tube expands due to the reason that the total internal pressure is greater than the external one. Such a perturbation also causes that the denser plasmas appear at around $x=-2$\,Mm, $x=0$\,Mm and $x=2$\,Mm, and the magnetic fields start to dip at the three locations. The $\exp$ function in the perturbation indicates that the highest plasma density appears at around $x=0$\,Mm, which causes the magnetic fields to sink most there and become U-shaped. When the U-shaped magnetic fields at around $x=0$\,Mm are continuelly pulled down by the heavy downflows, the generated magnetic pressure causes the horizontal flows to press the magnetic fields with opposite directions in the U-shaped part together and magnetic reconnection happens. The similar methods for triggering flux emergence as in this work are usually applied in the simulations which do not include the radiative cooling effect below the solar surface  \citep[e.g.,][]{Archontis2004,Syntelis2015}, which are very different from the previous RMHD simulations \citep[e.g., ][]{ Danilovic2017b}. The radiative transfer equations were solved in those RMHD simulations and the self-consistent convection zone was generated, then the flux emergence was triggered by convections.        

The periodical boundary conditions are applied in the $x$-direction. The outflow boundary conditions are applied in $z$-direction, which have been described clearly in the previous paper \citep{Ni2021}. The inflow and outflow boundary conditions are separately used at the bottom boundary and the up boundary in the $y$-direction. The fluid is only allowed to flow out of the simulation domain when the outflow boundary is applied. Conversely, the fluid is only allowed to flow into the simulation domain when the inflow boundary is applied.

\section{Numerical Results}

\subsection{The  formation of an Ellerman Bomb} \label{subsec:tables}

The previous section has described how the U-shaped magnetic fields are generated at around $x=0$\,M, a ball of heated plasmas similar to an EB is formed in this region after $t=260$\,s. Figure.3 shows the 3D plot of the EB area at $t=280$\,s. In this figure, one can see a yellow high-temperature ($\sim9000$\,K) strip shaped structure, which represents the EB's location that is entangled by magnetic field lines of a small twisted flux tube. Figure 4 (a), (b), and (c) show the temperature slices in the $x$-$y$ plane at $z=0$\,km at three different times, respectively. From these subgraphs, one can see that the flame-like high temperature region appears at around $x=0$\,km. The temperatures inside this event are between $5000$\,K and $10,000$\,K, it starts at the location approximately $500$\,km above the bottom of the photosphere and extends to $1100$\,km in the $y$-direction. Figure.4 (d), (e), (f) show the plasma $\beta$ distributions in the same planes as those in 4(a), (b) and (c), respectively. One can see that the plasma $\beta$ in most areas of the upper simulation domain is much larger than 1 because of the very weak magnetic field there. In the EB area, $\beta$ is also larger than 1. Figure.4 (g), (h), (i) show the the distributions of By in the $x$-$z$ plane at $y=0$\,km at the same three corresponding times as above, the black arrows represent  the velocities in the planes. The three subgraphs show that the magnetic fields with opposite directions are close to each other at around $x=0$\,km. However, there is no obvious magnetic cancellation in this plane during the formation process of the EB. The reason is that the emerging process continues to replenish magnetic fields during the magnetic reconnection process. Near the left and right boundaries of Figure 4.(a) (b) (c), there are two high-temperature areas where their temperatures are more than $20,000$\,K. We are not sure if they are caused by boundary conditions or other effects. 

\begin{figure}
	\begin{center}
		\includegraphics[width=14.0cm]{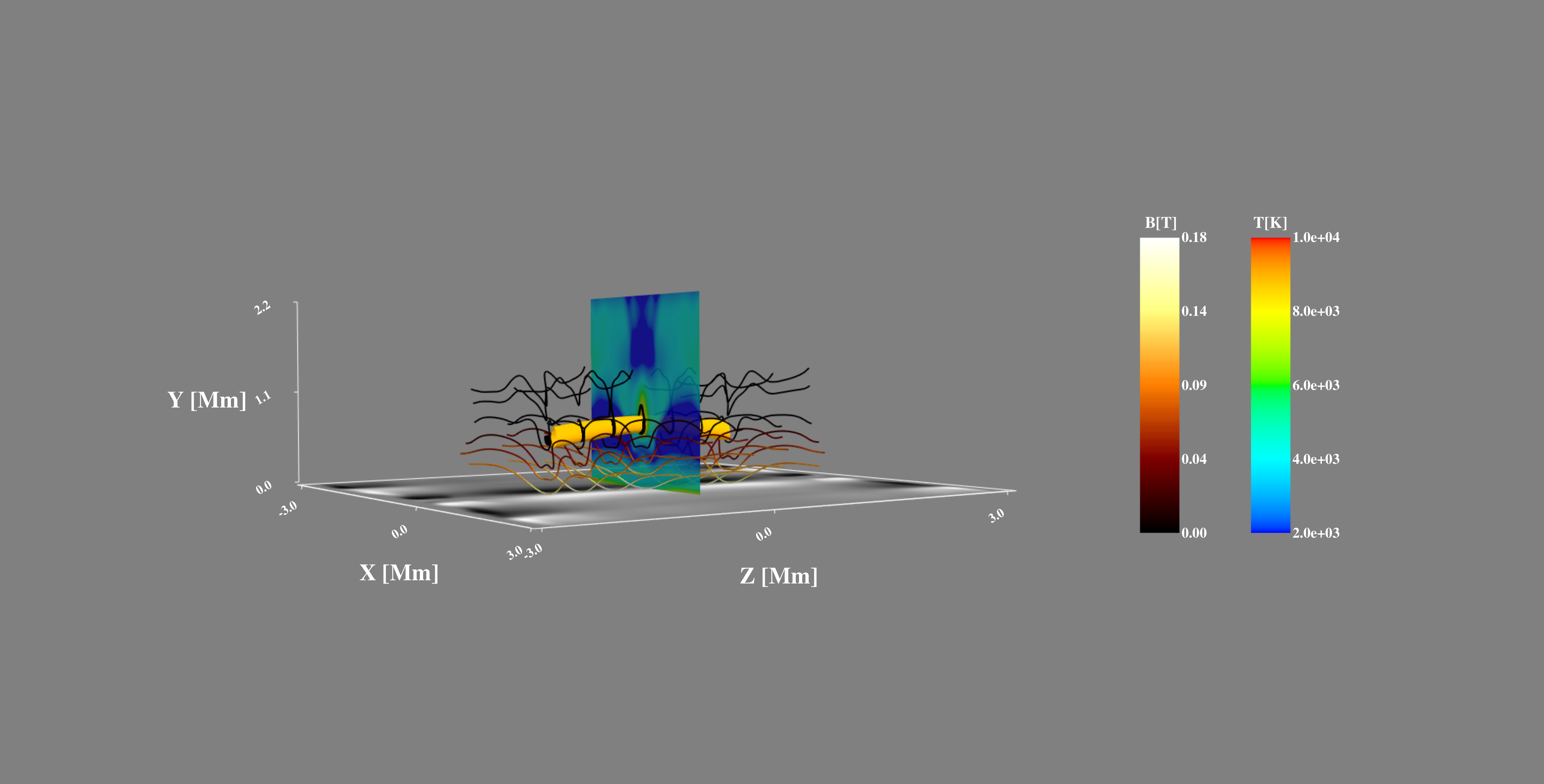}
		\includegraphics[width=14.0cm]{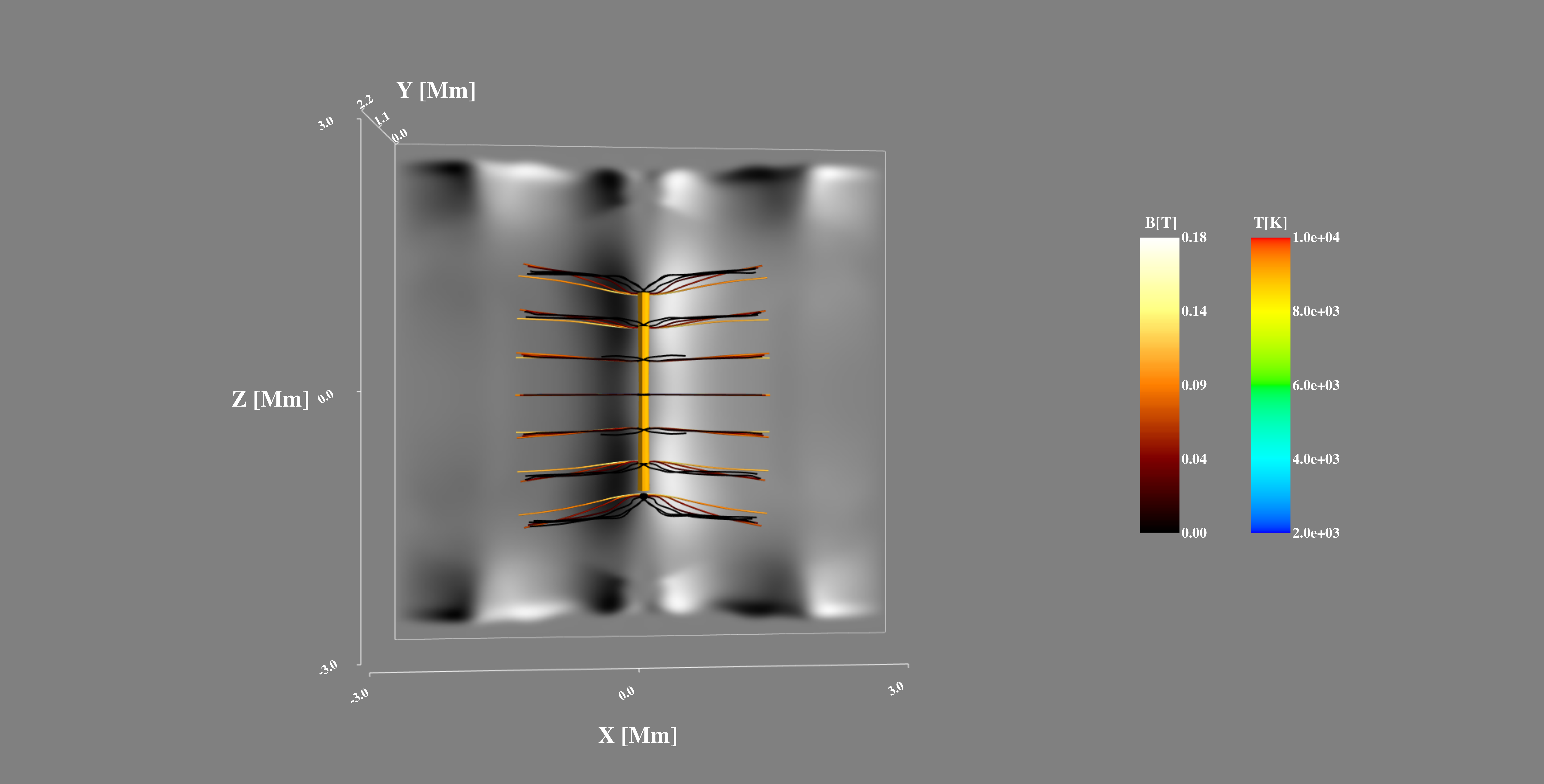}
	\end{center}
	\caption{The three-dimensional views of the EB area from two different viewing at $t=280$\,s. The bottom gray-scale slice shows the magnetic fields along the $y$-direction ($B_y$) in the $x$-$z$ plane at $y=0$\,km. The lines with colors represent the three-dimensional magnetic field lines, the different colors represent the different strengths. The yellow strip shaped structure represents the EB's location, where the temperature is about $9000$\,K. The vertical slice represents the temperature distribution in the $x$-$y$ plane at $z=0$\,km.}	
\end{figure}

\begin{figure}
	\begin{center}
		\includegraphics[width=0.8\textwidth]{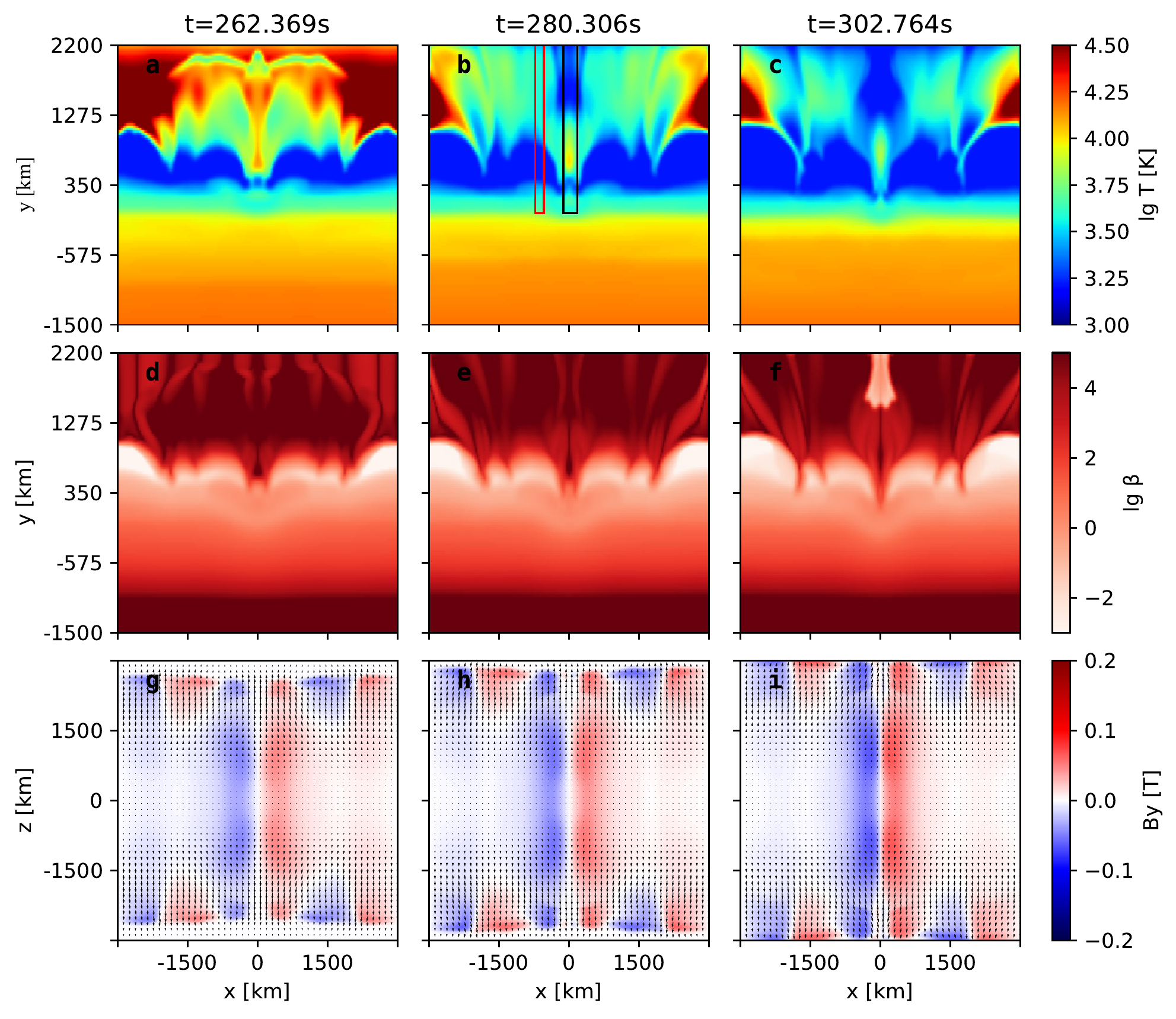}
	\end{center}
	\caption{The distributions of different variables in the 2D slice at three different times during the formation stage of the EB. (a), (b) and (c) show the distributions of the temperature in the $x$-$y$ plane at $z=0$\,km; (d), (e) and (f) show the distributions of the plasma $\beta$ in the $x$-$y$ plane at $z=0$\,km; (e), (g) and (h) show the distributions of the vertical magnetic field ($B_y$) in the $x$-$z$ plane at $y =0$\,km, the black arrows represent the velocity in this plane.The black rectangular box and the red rectangular box in (b) represent the regions for synthesizing the averaged H$\alpha$ spectral lines in the EB and in the nearby environment, respectively.}	
\end{figure}

In order to verify the observed event is an EB, the radiative transfer code MULTI 1.5D \citep{Carlsson1986} is applied to synthesize the H$\alpha$ line core and wing images and calculate the corresponding H$\alpha$ spectral line profile by using the data from our MHD simulations. Figure.5(a) and 5(b) show the synthesized H$\alpha$ core ($\lambda=6562.86$ \AA\ ) and wing ($\lambda=6563.83$ \AA\ ) images in the $x$-$z$ plane with the line of sight along the $y$-direction. Comparing Figure.5(a) and 5(b), one can see the region in the center around $x=0$\,km has strong emissions in H$\alpha$ wing image but the emission is not obvious in H$\alpha$ core image, which is consistent with the characteristics of an EB. However, the long strip shape with a length of $3$\,Mm has not been shown in the previous observations of EBs. The previous 3D RMHD simulations show that the shape of the synthesized H$\alpha$ wing image is different when the viewing angle is different \citep{Danilovic2017b}. The shape and size of the flame-like high temperature structure in the $x$-$y$ plane as shown in Figure.4 is more similar as the observed EBs. We can speculate that the synthesized H$\alpha$ wing image will be more like the observed one if the sight line could be more close to the $z$-direction. However, the MULTI 1.5 D code is not suitable for synthesizing the image and spectral line profile when the sight line is parallel to the solar surface (such as in the $z$-direction). Synthesizing H$\alpha$ wing images along a direction parallel to the solar surface is beyond this work. 
	
Figure.6(a) and (b) display the H$\alpha$ spectral line profiles calculated by the MULTI 1.5D. The blue line in Figure.6(a) is the synthesized H$\alpha$ spectral line profile along the $y$-direction by using the data passing through the EB area [inside the black rectangular box in Figure 4(b)], that solid red line represents the synthesized one by using the data passing through the area near the EB [inside the red rectangular box in Figure.4(b)]. The black solid profile in Figure.6(b) is derived by subtracting blue and red profiles in Figure.6(a). Figure.6(a) shows that the emission intensity in the H$\alpha$ wings from the EB area is much stronger than that from the nearby atmosphere, but the emission intensity in the H$\alpha$ core from the EB area is much weaker. Therefore, the substracted result in Figure.6(b) shows emissions in the H$\alpha$ wings but absorptions around the core center. Such a moustache-like structure in Figure. 6(b) is similar as the substracted H$\alpha$ line profile as shown in Figure.6(d) from observations \citep{Pariat2007}. In Figure.6(b), the maximum emission is at around H$\alpha$ $\pm$ 1.3 \AA\ and the emission is fading at $\pm$ 4.3\AA\ , which is consistent with the previous descriptions of the observed H$\alpha$ spectral line profile of the EB \citep{Severny1968}. However, comparing Figure.6(a) and 6(c), one can find some differences. The emission intensities of the synthesized ones are weaker than the observed ones, but the differences between the spectral line profiles of the EB area and the nearby atmosphere as shown in Figure. 6(a) are more significant than the observed ones, which then results that the maximum emission intensity in the wing is larger and the absorption in the core is deeper than the observed ones in the substracted H$\alpha$ spectral line profile. 
	
The unusual very cold plasmas with a temperature lower than $3000$\,K appear above and near the EB area as shown in Figure. 4, which is probably resulted from the over-simple radiative cooling model in this region or the up boundary conditions. As we know, the spectral line profiles include the integration effect of the plasma from the solar surface to the observation equipment. However, the maximum height of the atmosphere along the y-direction is only $2.2$\,Mm in our simulations. During the formation process of the EB, a part of the plasmas in the chromosphere is ejected out of the simulation domain. Therefore, the synthesized H$\alpha$ spectral line profiles lose the information of the ejected plasmas. The above reasons might cause the deviations between our synthesized spectral line profiles with the observational ones.

\begin{figure}
	\begin{center}
		\includegraphics[width=0.8\textwidth]{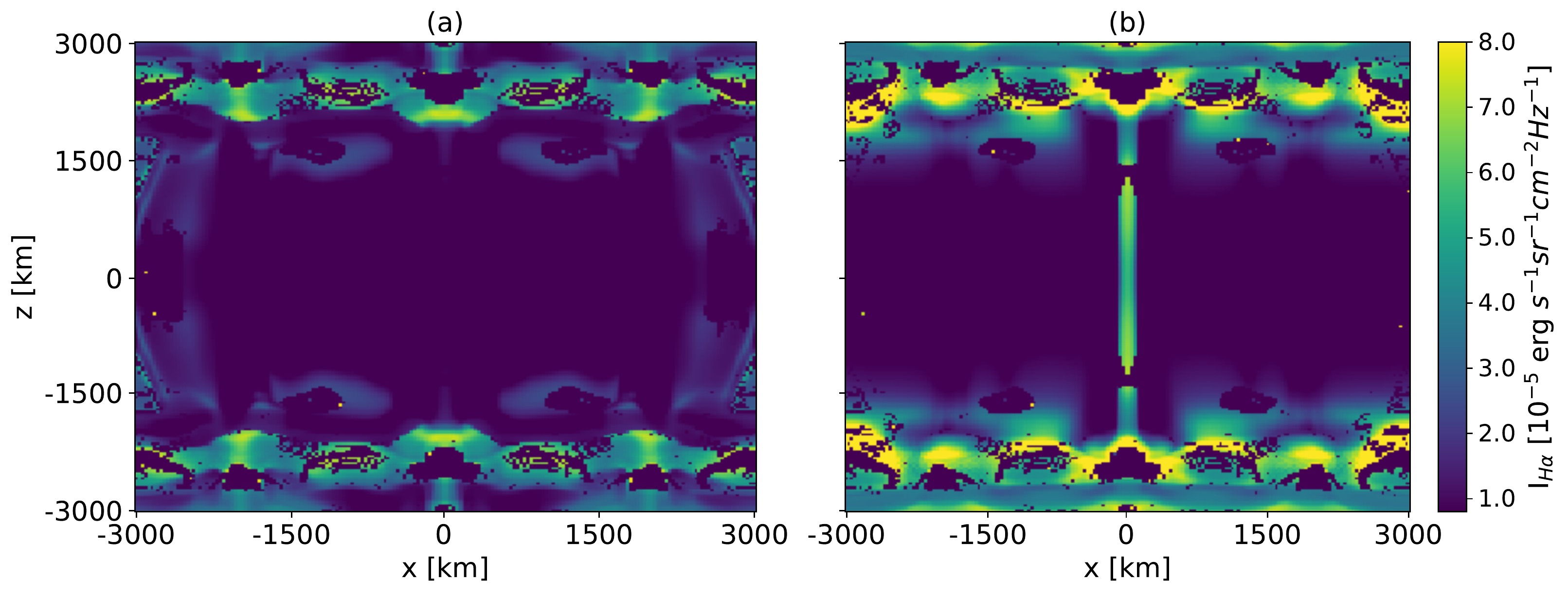}
	\end{center}
	\caption{(a) shows the synthesized H$\alpha$ core ($\lambda=6562.86$ \AA\ ) image and (b) shows the synthesized H$\alpha$ wing ($\lambda=6563.83$ \AA\ ) image in the $x$-$z$ plane. The line of sight is along the $y$-direction.}	
\end{figure}

\begin{figure}
	\centerline{\includegraphics[width=14.0cm, clip=]{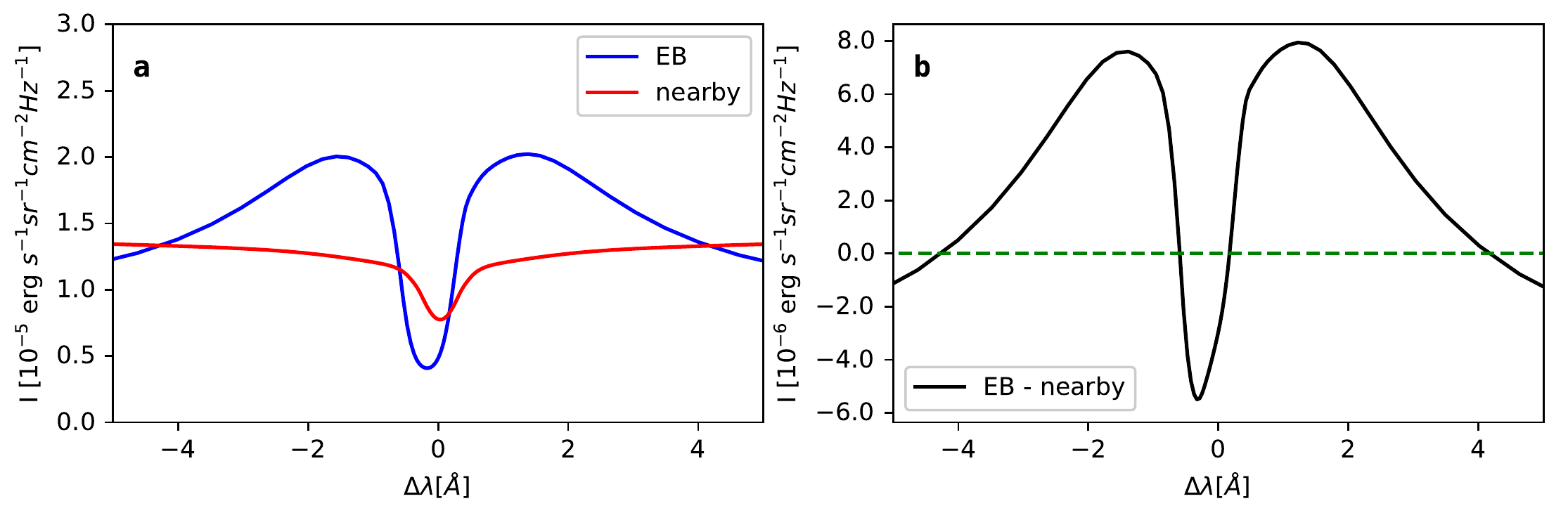}}
	\centerline{\includegraphics[width=14.0cm, clip=]{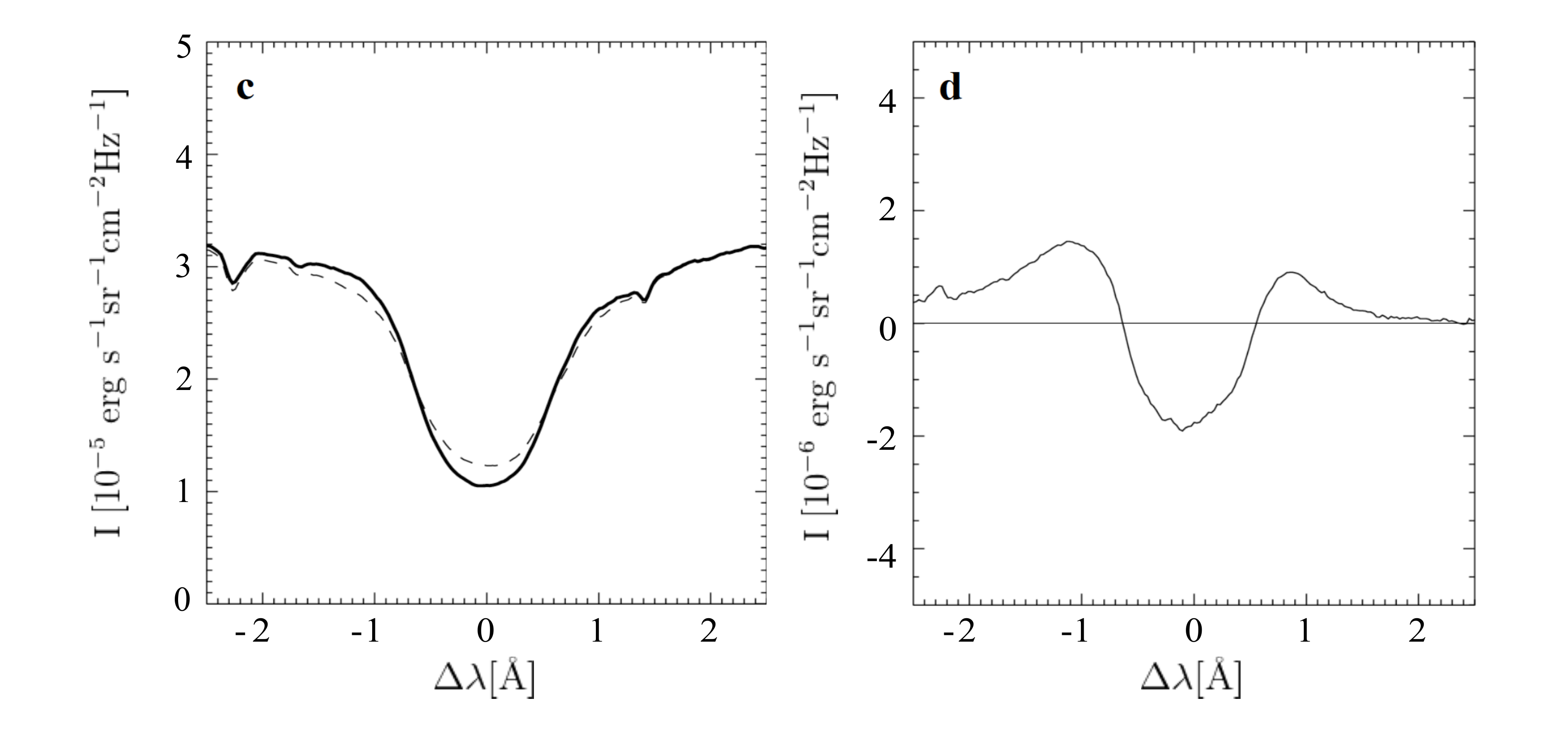}}
	\caption{The results of the synthesized H$\alpha$ spectral line profiles by using the MULTI 1.5D code. The blue line in (a) is the spectral line profile passing through the EB [the red rectangular box in Figure 4(b)], the red line in (a) is the spectral line profile passing through the nearby atmosphere [the red rectangular box in Figure 4(b)]. The black solid profile in (b) is the result by subtracting the data of the red profile from the data of the blue profile in (a). Panels (c) and (d) are the THEMIS/MTR H$\alpha$ spectral line profiles of one EB, the bold line in (c) is the EB profile and the dashed line is the nearby reference H$\alpha$ profile, (d) is the result by subtracting the nearby reference profile from the EB profile. The images in (c) and (d) reproduced with permission from \cite{Pariat2007}, copyright by ESO. }	
\end{figure}

In the following two paragraphs, we will study the formation mechanisms of the EB. Figure.7 shows the distributions of different physical variables in the zoomed in area with the EB in the $x$-$y$ plane at $z=0$\,km, $t=280$\,s. The distributions of the velocity (the thin white arrows) and plasma temperature (the color contour maps) are presented in Figure.7(a). The highest temperature in the EB is about $10,000$\,K. In $x$-direction, the plasmas on two wings of that EB move toward each other. Figure.7(b) shows the distributions of the current density in the $z$-direction ($J_z$) and the two-dimensional magnetic field lines. One can see the U-shaped magnetic fields above the photosphere and a short reconnection current sheet inside the black rectangular box. Comparing Figure.7(a) and 7(b), one can find that the current sheet is below the EB and about $200$\,km away from the EB, and there is no obvious high temperature plasmas inside the curremt sheet. Figure.8 displays the vertical velocity ($V_y$) along the thick vertical white arrow in Figure.7(a), the two blue dashed lines represent the locations of the two ends of the current sheet, and the two red dashed lines represent the locations of the two ends of the EB. One can see that the current sheet is more than 150 km away from the EB area and the plasma velocity inside the current sheet is downward, which indicates that neither the magnetic reconnection process nor the reconnection outflows can heat the plasmas in the EB area. 

The distributions of the divergence of the velocity and the plasma density are displayed in Figure.7(c) and 7(d), respectively. One could see that two obvious blue dividing lines representing the areas with large values of $\nabla \cdot \textbf{v}$ are located on both sides of the EB region in Figure.7(c), which indicate the possible shock fronts. Figure.9 shows the distributions of different variables along the thick white arrow in Figure.7(c). The shock fronts which are passed through by the thick white arrow can be considered along the $y$-direction, and the variables along the $y$-direction are approximately parallel to the shock front. The black solid line and the red dotted line in Figure.9(a) represent the parallel magnetic field ($B_y$) and vertical one ($B_x$) respectively. Figure.9(b) shows the distributions of the vertical velocity ($V_x$), the parallel velocity ($V_y$) and the Mach number $Ma$, $Ma = V_x / \sqrt{V_{\rm s}^{2}+V_{\rm Ay}^{2}}$, where $V_{\rm s}$ is the sound speed and $V_{\rm Ay}$ is the Alvfen speed calculated by using magnetic field in y direction. Figure.9(c) and 9(d) show the logarithm of the plasma density and temperature respectively. Crossing the left shock front (the green dashed line in Figure.9), the parallel component of the magnetic field ($B_y$), density, and temperature are all increased, and the vertical velocity $V_x$ is decreased. We can also see the Mach number is larger than 1 ahead the shock and decreases to a value smaller than 1 behind the shock front. All these characteristics indicate that the thick white arrow in Figure.7(c) passes through two fast-mode shocks, which heat the plasmas in the EB area. We should point out that the low resolution in this work smooths out the sharp structures around the shock fronts. 

\begin{figure}
	\begin{center}
		\includegraphics[width=0.8\textwidth]{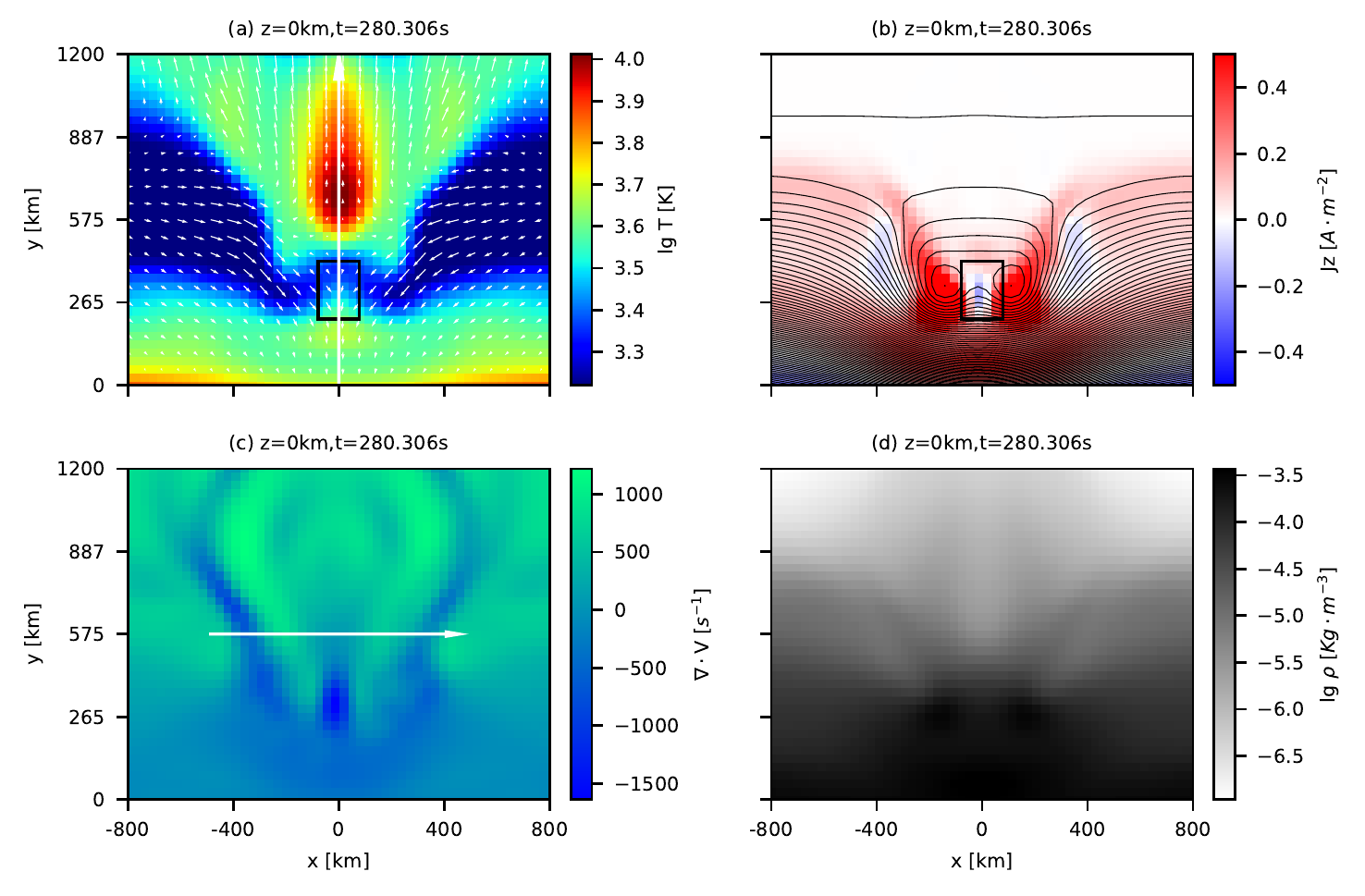}
	\end{center}
	\caption{The distributions of different variables in the $x$-$y$ plane at $z=0$\,km at $t=280$\,s. (a) shows the distributions of velocity (the white thin arrows) and temperature;  (b) shows the distributions of current density in the $z$-direction ($J_z$) and the two-dimensional magnetic field lines, the rectangular box in (b) is the area where the reconnected current sheet is located; (c) shows the distributions of the divergence of the velocity in this plane; (d) is the density distribution.}	
\end{figure}

\begin{figure}
	\begin{center}
		\includegraphics[width=0.6\textwidth]{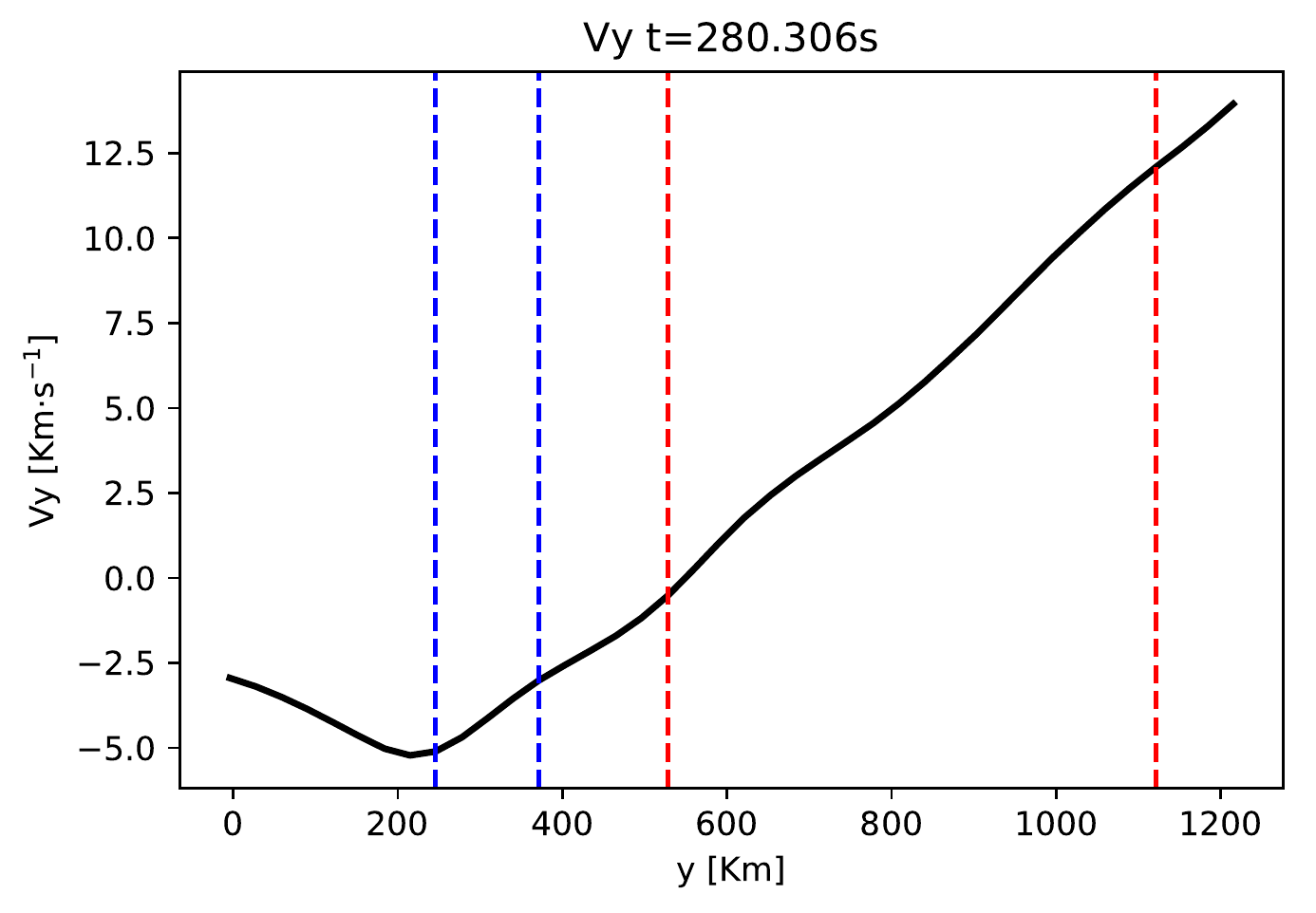}
	\end{center}
	\caption{The distribution of the vertical velocity ($V_y$) along the thick white arrow in Figure 7(a). The blue dashed lines represent the two ends of the current sheet, and the red dashed lines represent the two ends of the EB.}	
\end{figure}

\begin{figure}
	\begin{center}
		\includegraphics[width=0.8\textwidth]{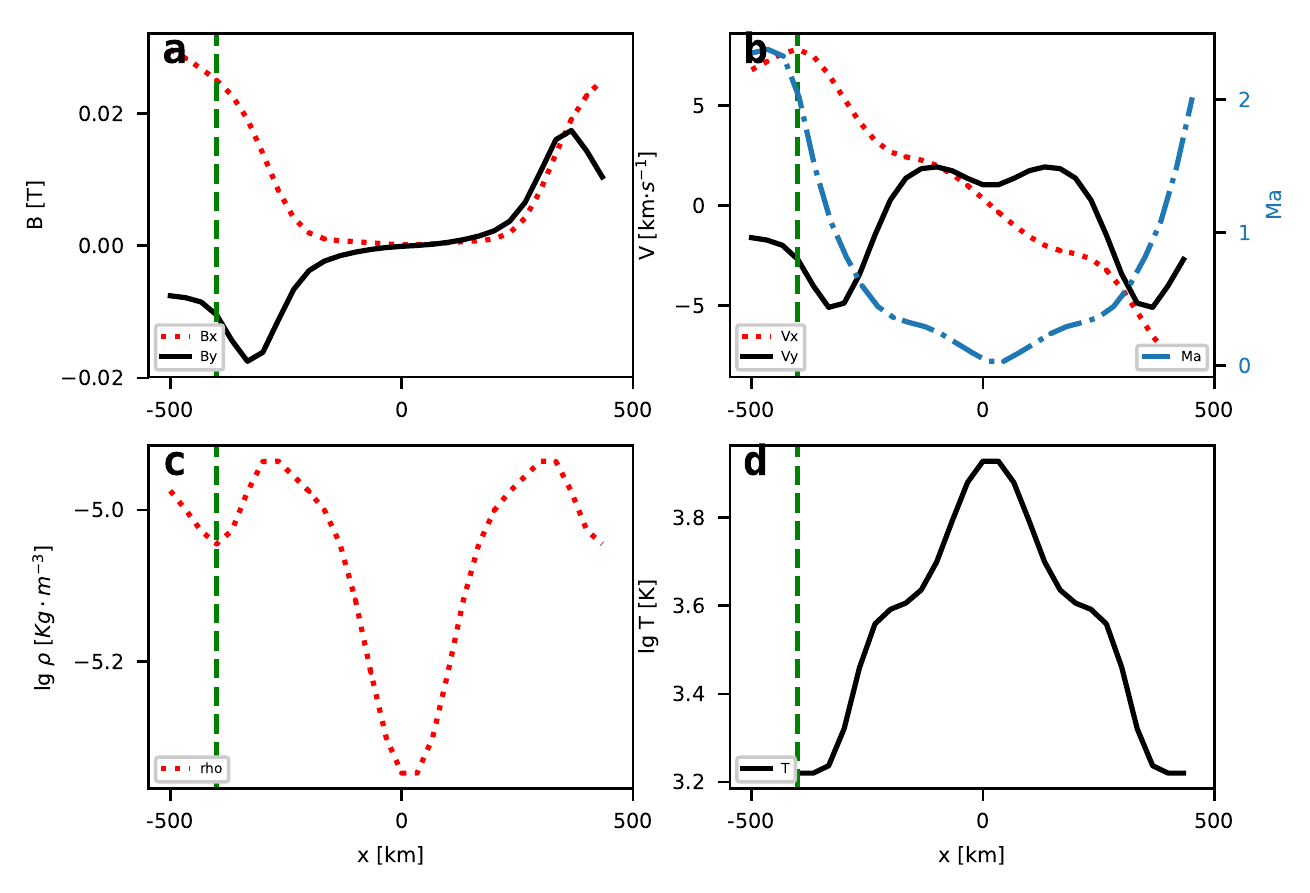}
	\end{center}
	\caption{The changes of different variables along the thick white arrow in Figure 7(c), the green dashed line represents the left shock front. The black solid line and the red dotted line represent the $B_y$ and $B_x$ in (a) respectively; $V_x$, $V_y$ and the Mach number $Ma$ are shown in (b); the logarithm of the plasma density is shown in (c); the logarithm of the temperature is shown in (d).}	
\end{figure}

The whole coherent scene of the EB formation process is very clear by analyzing the time dependent simulation results. The Parker instability triggered by initial perturbations leads to the emerging of the magnetic fields and the formation of U-shaped magnetic fields. The special initial perturbation formula makes that the magnetic fields to sink most around the center of the simulation domain. When the U-shaped magnetic fields in this region around $x=0$\,km are continuouly pulled down by the heavy downflows, the generated magnetic pressure causes the horizontal flows to press the magnetic fields with opposite directions in the U-shaped part together and magnetic reconnection happens. However, the short reconnection site during the formation process of the EB is located at a very low altitude in the photosphere ($\sim250-400$\,km above the solar surface). After the newly formed twisted flux rope during the magnetic reconnection process floats to a higher location around the solar TMR,  it is then heated by the shocks driven by the strong horizontal flows with opposite directions at the both wings of the U-morpha magnetic fields. Figure.3 clearly shows that the hot plasmas in the EB are located inside the newly formed twisted flux rope, which is corresponding to the magnetic island in the 2D simulations. Such a scenario is consistent with the previous 2D results that the heated plasmas in the reconnection region are mostly located inside the magnetic islands. The magnetic island corresponding to the newly formed twisted magnetic flux rope can not be seen in Figure.7(b). Since the strength of magnetic fields in the twisted magnetic flux rope region is much weaker than that in the bottom region, the contour lines representing the magnetic field lines can not appear in the corresponding region in Figure.7(b).

The maximum temperature in the EB reaches around  $10,000$\,K in this simulation, which is a little bit higher than the maximum temperature of $9,000$\,K in the previous 3D simulation of the EB in an active region \citep{Danilovic2017b}. Figure. 7 and Figure. 9(c) show that the plasma density in the EB region is about $10^{21}$ which is consistent with the result of \cite{Danilovic2017b}. The previous 3D RMHD simulations showed that EBs are below the solar TMR, but the EB in our simulation extends from $\sim500$\,km to $\sim1100$\,km above the solar surface. The high density photospheric plasmas are ejected to the upper atmosphere probably by a stronger flux emerging process, which then results in a higher formation height in this work. The particular initial perturbation formula makes that the reconnection regions in U-shaped part of the magnetic fields are only located around $x=0$\,km and the variables do not change very much in $z$-direction during the reconnection process, which makes the simulations are more close to a 2.5 D MHD simulation. Therefore, the long strip-like structure of the EB is formed in the z-direction, which are usually not shown in the observational ones. The previous simulations include a more realistic convection zone under the photosphere \citep[e.g.,][]{Danilovic2017a, Danilovic2017b, Hansteen2017, Hansteen2019}, the convections make the variables change strongly with time in all the three directions, many U-shaped magnetic fields and reconnection regions then asymmetrically appear in different areas in the photosphere. Therefore, the long strip-like structure is not shown in those simulations. The EB was directly located in the reconnection region and heated by magnetic reconnection process in all the previous simulations. Though the small twisted flux rope with the EB is originally formed in the magnetic reconnection process, the EB is far away from the reconnection region and heated by the shocks as described above. Hence, we propose a new mechanism for the generation of the EB. 

\subsection{The formation of an UV burst} \label{subsec:tables}

The EB lasts about one minute and then cool down gradually until it disappears. The U-shaped part of the magnetic fields around $x=0$\,km continues to move downward due to the Parker instability, while the tops of the $\Omega$-shaped part on both sides of the U-shaped part continue to rise to the higher atmosphere, reaching the middle and up chromosphere. The reconnection region then extends from the photosphere to the up chromosphere. After $t=700$\,s, the plasmas with a high temperature of tens of thousands Kelvin appear in the reconnection region in the chromosphere. The following analyses prove that these high temperature plasmas are corresponding to the UV burst. Figure.10 shows the three-dimensional displays of the UV burst regions from two different viewing angles. The horizontal slice at $y=0$ represents the vertical magnetic fields at the solar surface, the vertical slice shows the distributions of temperature at $z=600$\,km, the colorful lines represent the magnetic field lines with different strength, the three-dimensional isosurface that represents the temperature is also plotted in Figure.10. One can find that two obvious high temperature regions are respectively located around $z=600$\,km and $z=-600$\,km, the maximum temperature reaches about $95,000$\,K. The high temperature plasmas are also entangled by magnetic field lines of a newly formed twisted flux tube during magnetic reconnection process. The temperatures and length scale of the high temperature plasmas are consistent with the characteristics of the observed UV bursts.  

\begin{figure}
	\begin{center}
		\includegraphics[width=14.0cm]{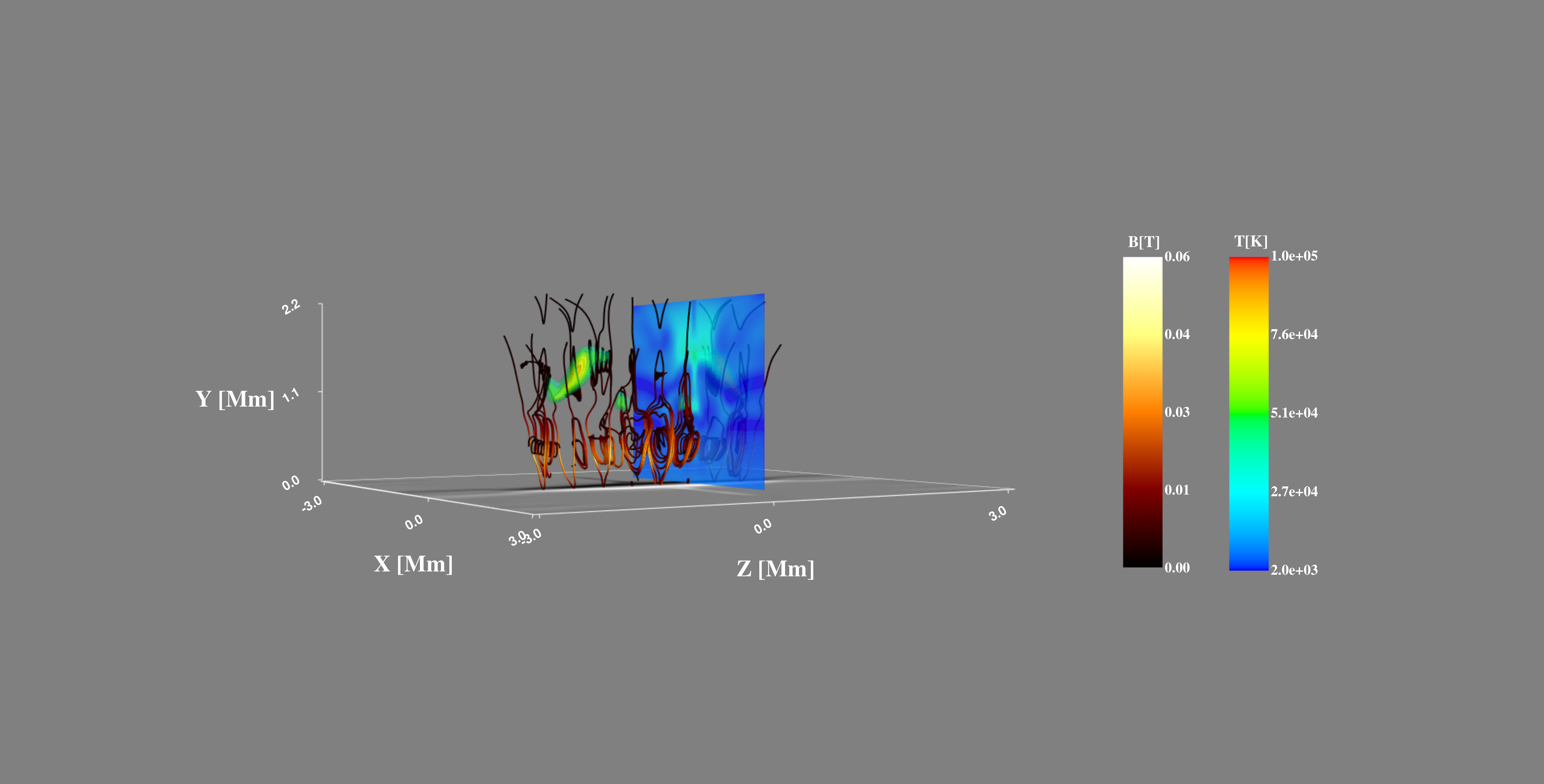}
		\includegraphics[width=14.0cm]{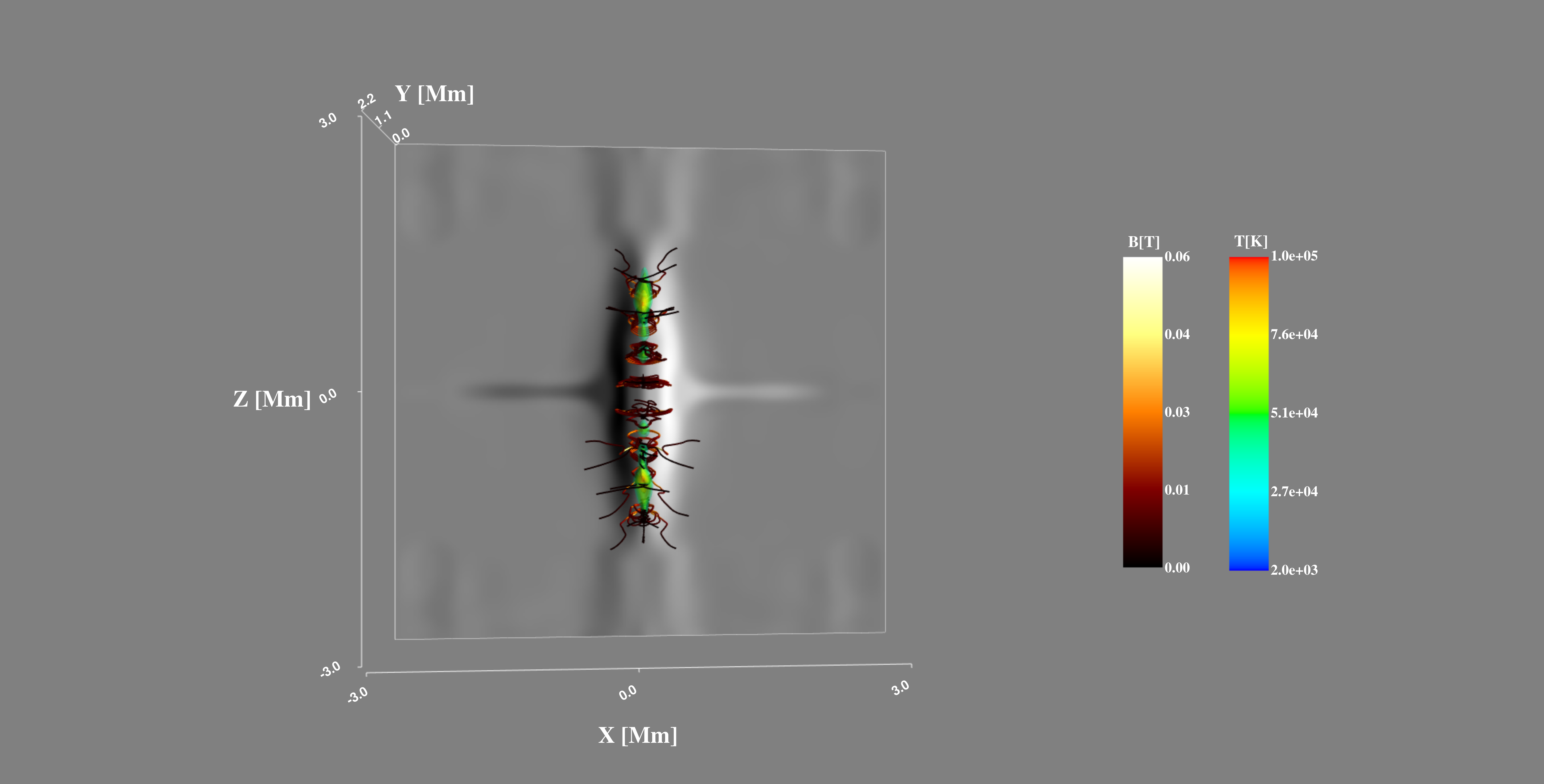}
	\end{center}
	\caption{The three-dimensional views of the UV burst area from two different angles at $t=732$\,s. The bottom gray-scale slice shows the magnetic fields along the  $y$-direction ($B_y$) in the $x$-$z$ plane at $y=0$\,km. The lines with colors represent the three-dimensional magnetic field lines, the different colors represnet the different strengths. The colorful three-dimensional isosurface represents the high temperature plasmas, the maximum tempersture is about $95,000$\,K. The vertical slice represents the distribution of the temperature in the $x$-$y$ plane at $z=600$\,km. }	
\end{figure}

Figure.11 displays the slices of the temperature and plasma $\beta$ in the $x$-$y$ plane at $z=600$\,km, and the vertical magnetic field distribution in the $x$-$z$ plane at $y=0$\,km, at three different times. The temperature distributions are displayed in Figure.11 (a), (b) and (c). Figure.11(b) indicates that the high temperature plasmas with a size of about one arcsec appear around $x=0$\,Mm in the chromosphere, the maximum temperature achieves approximately $40,000$\,K (the maximum temperature is different in different slices). Figure.11 (d), (e) and (f) show the distributions of plasma $\beta$, one can see that the plasma $\beta$ in the chromosphere is much smaller than that during the previous EB stage by comparing the results in these figures with those in Figure.4 (d), (e), (f). The gradually increased magnetic fields in the chromosphere by flux emerging process is the reason to cause the differences. The plasma $\beta$ around the inflow regions of the vertical reconnection current sheet is smaller than $1$, which agrees with the range of the plasma $\beta$ for generating the UV bursts in the previous 2D simulations \citep[e.g.,][]{Ni2016,Peter2019}. However, the plasma $\beta$ in the center of the current sheet around $x=0$\,km is still larger than 1, the annihilation of magnetic fields in the $x$-$y$ plane by magnetic reconnection and the very weak guide field in the $z$-direction cause the large plasma $\beta$ in this region. Such a scenario is also similar with the previous 2D numerical experiments \citep[e.g.,][]{Ni2015}. Figure.11 (g), (h), (i) show the distributions of the vertical magnetic fields and the velocity in the $x$-$z$ plane at $y=0$\,km, the magnetic cancellation is clearly observed in the region where the opposite magnetic fields meet with each other during the formation process of the UV burst. 

\begin{figure}
	\begin{center}
		\includegraphics[width=0.8\textwidth]{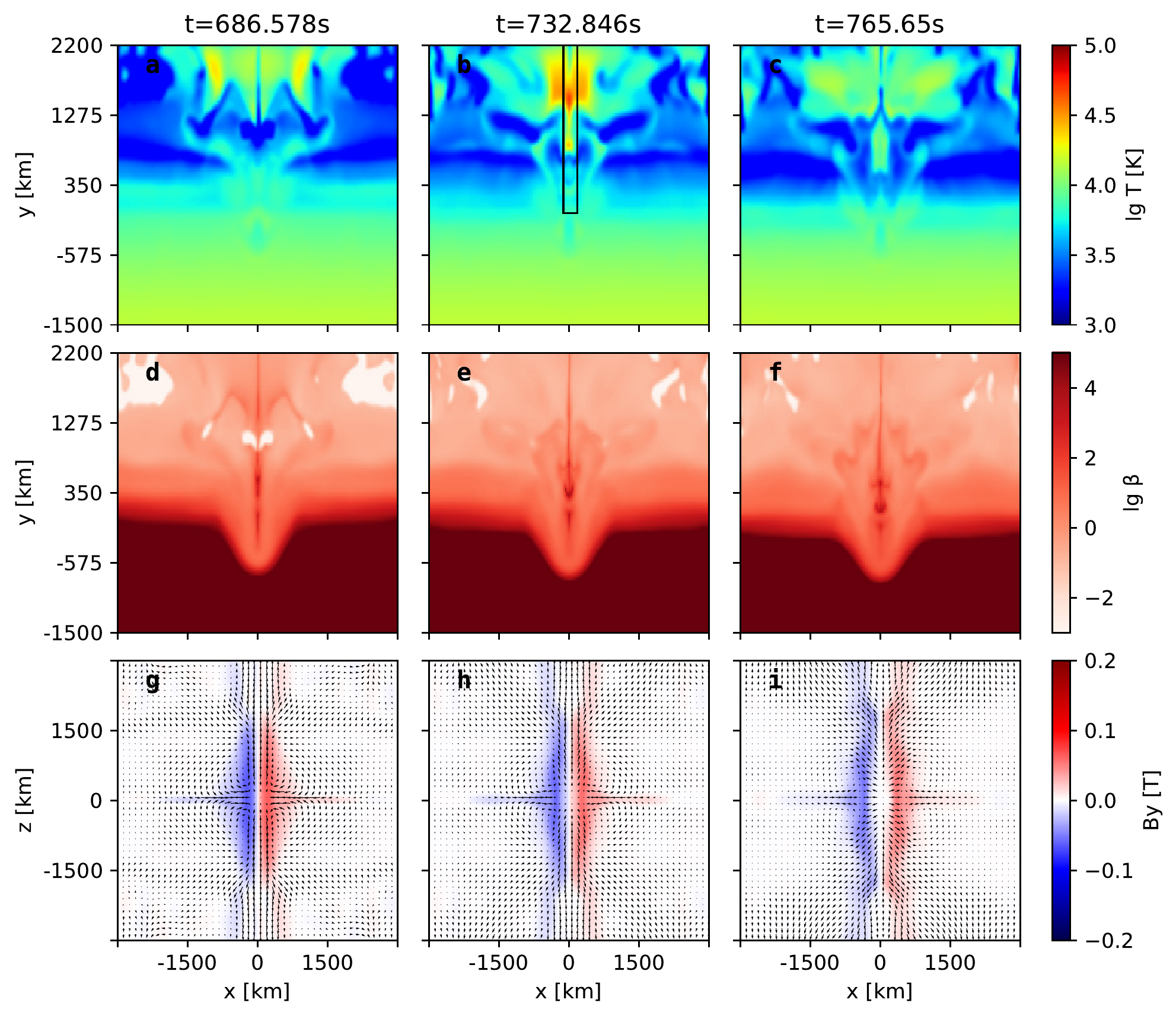}
	\end{center}
	\caption{The distributions of different variables in the 2D slice at three different times. (a), (b) and (c) display the temperature distributions in the $x$-$y$ plane at $z=600$\,km; (d), (e) and (f) displays the map of the plasma $\beta$ in the $x$-$y$ plane at $z=600$\,km; (e), (g) and (h) show the distributions of the vertical magnetic field ($B_y$) in the $x$-$z$ plane at $y=0$\,km, the black arrows represent the velocity in this plane. }	
\end{figure}

In order to explore this UV burst's radiation characteristics in the Si IV band, we have used the optically thin approximation method of spectral line synthesis and the atomic data package CHIANTI (version 9) to obtain the Si IV line profile and emission intensity images from different viewing angle \citep[e.g.,][]{Ni2021, Li2019, Innes2015, Dere2019}.  The total Si IV emission intensity in Figure. 12(a) and 12(b)  is calculated as
	\begin{eqnarray}
	I_{\rm Si\ IV}^{\rm tot} =  \int_{\rm s}  n_{\rm e} n_{\rm H} g(T) ds, 
	\end{eqnarray}
	where $n_{\rm e}$ is the number density of the electrons and $n_H$ is the one of protons, $g(T)$ represents the contribution function. The Si IV spectral line profile in Figure.12(c) is calculated as
	\begin{eqnarray}
	I_{\rm Si\ IV} =  \int_{\rm s} \phi_{\rm \lambda} n_{\rm e} n_{\rm H} g(T) ds, 
	\end{eqnarray}
	where $\phi_{\rm \lambda}$ is the relative velocity distribution function and is given by
	\begin{eqnarray}
	\phi_{\rm \lambda} = \frac{1}{\pi^{1/2} \Delta \lambda_{\rm D}} \exp \left[ -\left( \frac{\Delta \lambda +\lambda_0 \frac{v_{\rm s}}{c}}{\Delta \lambda_{\rm D}}  \right)^2 \right],
	\end{eqnarray}
	where $\Delta \lambda = \lambda - \lambda_0$ is the offset from the rest wavelength $\lambda_0=1393.755$ \AA\  and $v_s$ is the flow speed projected along the sight line. The expression of the thermal broadening $\Delta \lambda_D$ can be given by
	\begin{eqnarray}
	\Delta \lambda_D = \frac{\lambda_0}{c} \sqrt{\frac{2k_B T}{m_{\rm Si}}},
	\end{eqnarray}
	where $m_{\rm Si}$, $c$, $k_B$ and $T$ represent the atomic mass of silicon, the speed of light, the Boltzman constant and the temperature,  respectively.

The optical thin approximation makes the images can be easily synthesized from different directions. Figure.12(a) shows the synthesized image in the $x$-$y$ plane and the line of sight is along the $z$-direction. We find that the UV burst starts with a location at around $y=750$\,km and ends at around $y=1500$\,km in the $y$-direction. Figure.12(b) shows the synthesized image when the line of sight is at an angle of 45$^\circ$ from both the $x$ and $x=z$ directions. One can see that the strong UV emissions appear at four different locations in Figure.12(b), which is very different from the shape in Figure.12(a). The averaged Si IV spectral line profile in Figure.12(c) is calculated by using the data in the black rectangular box in Figure.11(b). The abscissa is the Doppler frequency shift velocity, and the ordinate is the emission intensity. The maximum Doppler frequency shift speed is about $20$\,km s$^{-1}$, which is consistent with the results of narrow-line-width UV bursts\citep[e.g.,][]{Hou2016}. As shown in Figure.12(a) and 12(b), the maximum total intensity is about $10^7$ erg s$^{-1}$ sr$^{-1}$ cm$^{-2}$. The maximum intensity in Figure. 12(c) is about $6.5\times10^5$ erg s$^{-1}$ sr$^{-1}$ cm$^{-2}$ \AA\ $^{-1}$. In the previous review paper \citep{Young2018}, the authors have shown both the images and the spectral line profiles of several UV bursts. Comparing the synthesized results from our simulations and the results about four observed UV burst in Fig. 2 in \cite{Young2018}, one can see that the maximum emission intensity in Figure.12 in this work is close to the observed ones.

\begin{figure}
	\begin{center}
		\includegraphics[width=0.9\textwidth]{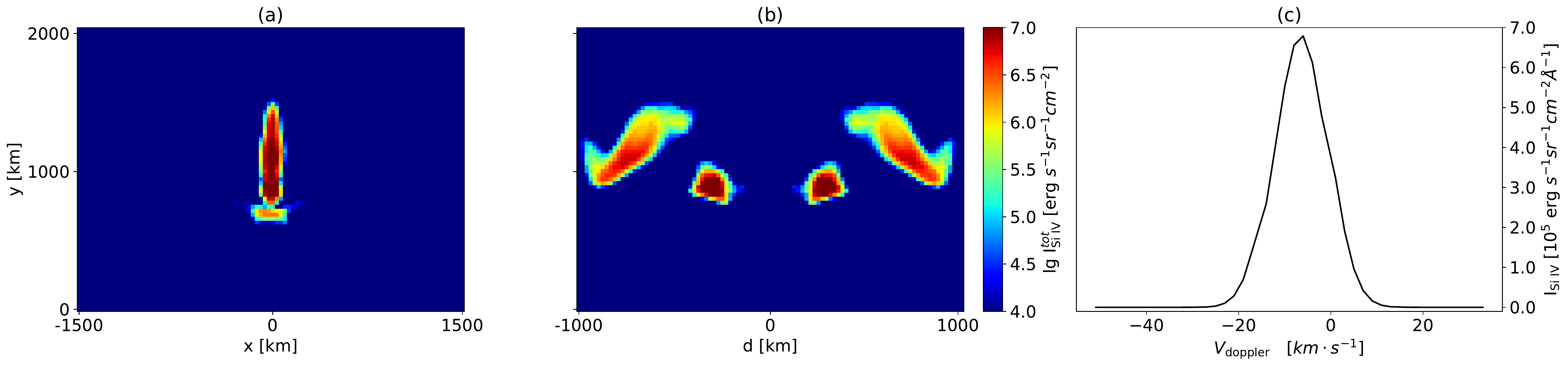}
	\end{center}
	\caption{(a) shows the synthesized Si IV emission image in the $x$-$y$ plane and the line of sight is along the $z$-direction.(b) shows the synthesized Si IV emission image when the line of sight is at an angle of 45 from both the $x$ and $x=z$ directions. (c) shows the average synthesized Si IV spectral line profile in the UV burst region near $x=0$.}	
\end{figure}

Figure.13 shows the distributions of different variables in the $x$-$y$ plane at $z=600$\,km at $t=732$\,s. In Figure.13(a), the white arrows represent the distributions of the velocity and the color contours represent the distributions of temperature. Figure.13(b) shows the distributions of the current density and the two-dimensional magnetic field lines, one can see that the region with strong Si IV emission intensity as shown in Figure.12(a) is exactly located in the reconnection current sheet region. From Figure.13(b), one can also see that plasmoids are generated in the reconnection region. Since the magnetic fields with opposite directions are pushed to approach each other by strong horizontal flows with opposite directions, and the velocity of horizontal inflows is even larger than the vertical outflow velocity, we can conclude that it is a driven magnetic reconnection process. Such a driven magnetic reconnection process is different from the previous ones triggered by the small initial perturbations \citep[e.g.,][]{Ni2015,Ni2016,Ni2018a}. The plasmas might also be heated by the strong compressing in such a process.

\begin{figure}
	\begin{center}
		\includegraphics[width=0.8\textwidth]{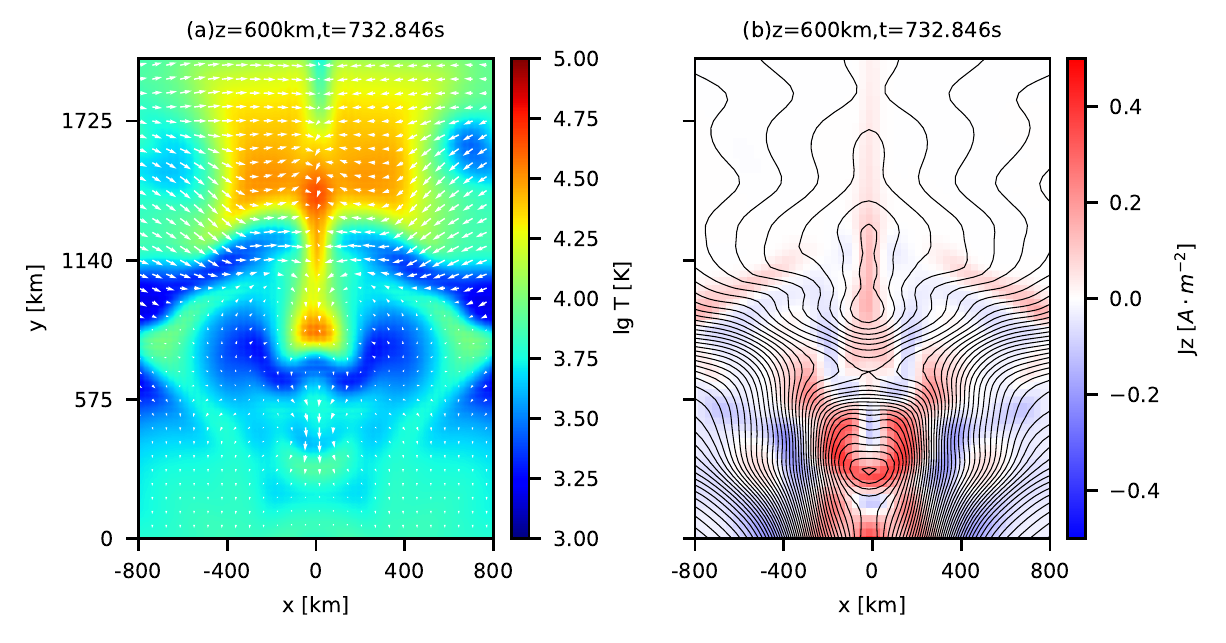}
	\end{center}
	\caption{The distributions of different variables in the $x$-$y$ plane at $z=600$\,km at $t=732.846$\,s. (a) shows the distributions of velocity (the white arrows) and temperature; (b) shows the distributions of the current density in the $z$-direction ($J_z$) and the two-dimensional magnetic field lines.}	
\end{figure}

The previous numerical results \citep{Hansteen2019} show that a long vertical current sheet extend from the photosphere to the transition region is formed during the flux emerging process, and the lower cool part and the hot upper part of this current sheet correspond to the EB and UV burst, respectively. In this work, we also find the vertical long current sheet that can be seen in Figure.13. In the vicinity of the high temperature UV burst, the colder plasmas with a temperature lower than $10,000$\,K also appear. We want to know whether these regions can produce the observed characteristics of EBs, and whether they can appear at about the same height as the UV burst. Therefore, we use the radiation transfer code to calculate the H$\alpha$ spectral line profiles in these regions by integrating corresponding values along the $y$-direction from $y=600$\,km to $y=2200$\,km. We find that only a few percent of the integrating results in these regions show the spectral line profiles that are similar as the observed EBs. We also tried to synthesize the emission intensity map at the H$\alpha$ wing with $\lambda$=6563.83 \AA. However, there is no obvious brightening in these regions in the synthesized image, no matter the integrating is from $y= 600$\,km to $y= 2200$\,km or from $y=0$\,km to $y=2200$\,km. As discussed in the above subsection, the losing of the chromospheric plasmas and an over-simple radiative cooling model might be the reasons to cause such a result. We can not conclude if the UV burst relates to EBs or not. 

Comparing with some of the previous simulations \citep{Rouppe2017, Hansteen2017, Hansteen2019}, the UV burst in this work is located at a lower altitude above the solar photosphere and the density in the UV burst region ($\sim10^{19}-10^{20}$\,m$^{-3}$) is higher. However, such a formation height and plasma density in the UV burst is similar as the recent 2D simulations \citep{Ni2021}. The synthesized maximum Si IV emission intensity is close to the observational ones \cite{Young2018}. However, the width of the spectral line profile is only about $20$\,km s$^{-1}$, which is different from the UV burst with a wide line profile of about $100$\,km s$^{-1}$, but similar with the narrow-line-width UV bursts. As discussed in the previous paper \citep{Ni2021}, the reconnection outflow velocity and the direction of the line of sight are crucial to the width of the Si IV spectrum line profile. Since the reconnection magnetic fields are not  strong enough in this work, the reconnection outflow velocity is not high enough to generate the wide-line-width UV burst. The numerical resolution in this work is similar as some of the previous 3D simulations \cite{Hansteen2017, Hansteen2019}. However, the twisted magnetic flux rope generated in magnetic reconnection process is not shown in those papers, but the non-Gaussian line profile of Si IV might indicate that there were the small magnetic flux ropes in the reconnection regions in those simulations. For the first time, we clearly show that the small twisted magnetic flux ropes (corresponding to the magnetic islands in the 2D plane) are generated during the magnetic reconnection process in the solar chromosphere. Our numerical results also show that the hot UV burst regions are located in the interior of the twisted flux rope. As mentioned in the above subsection, such a scenario is close to the previous 2D simulations that the heated plasmas in the reconnection region are mostly located inside the magnetic islands.

\section{Summary and discussion}

In this work, the emerging process of a single untwisted flux rope is studied based on 3D MHD simulations. Due to the special density destabilization at t=0 s in our simulation, the plasma environment is unstable and then the initial magnetic tube expands and emerges above the photosphere under the influence of the Parker instability. The U-shaped magnetic fields appear around $x=0$\,km and move downward, while the tops of the $\Omega$-shape part on both sides of the U-shaped part continue to rise to the higher atmosphere. Magnetic reconnection happens inside the U-shape part of the magnetic fields, and the current sheet gradually emerges at a higher location. The EB and UV burst are generated successively in this emerging process. The radiative transfer code Multi 1.5D is applied to synthesize the H$\alpha$ spectrum line profile of the EB. The Si IV emissivity and spectral line profile of the UV bursts have also been synthesized. We have analyzed and discussed the mechanisms for generating the EB and UV burst, respectively

The EB starts to appear earlier and lasts for about 80 seconds. After the new twisted flux rope generated in the reconnection region in the photosphere rises to a greater altitude in the low chromosphere, it is then heated by the shocks driven by the strong horizontal flows with opposite directions at the both wings of the U-morpha magnetic fields. The synthesized H$\alpha$ spectral line profile shows a mustache-like structure, and the maximum emission is at around H$\alpha$ $\pm$ 1.3 \AA\ and the emission is fading at $\pm$ 4.3\AA\ , which are consistent with the previous descriptions of the observed H$\alpha$ spectral line profile of the EB \citep{Severny1968}. However, the deviations between the simulations and observations still exist as described in Section 3.1. The EB-like event extends from $\sim500$\,km to $\sim1100$\,km above the solar surface. The UV burst-like event starts to appear six minutes later after the EB disappears and it lasts for about 60 seconds.  The driven reconnection process triggered by the strong inflows at two sides of the U-shaped magnetic field structures is the primary mechanism to result in the plasm heating in the UV burst. The maximum synthesized Si IV emission intensity is $\sim10^6$ erg s$^{-1}$ sr$^{-1}$ cm$^{-2}$ \AA\ $^{-1}$ and  it is close to the observed ones presented by \cite{Young2018}. The line width of the Si IV profile is approximately equal to $20$\,km s$^{-1}$ and it is only similar to those narrow-line-width UV bursts. The UV burst starts with a location at around $y=750$\,km and ends at around $y=1500$\,km in the $y$-direction.     
	
	The main conclusions of this work are presented as follows: 
	
	1. Though the resolution in this work is much lower than that in the previous 2D simulations, the small newly formed twisted flux ropes (corresponding to the magnetic islands in 2D simulations) are still found in the reconnection process and both the EB and the UV burst are located in these twisted flux ropes, which were not shown in the previous 3D simulations of UV bursts or EBs. These results also prove that the unstable magnetic reconnection process with plasmoid instability indeed can appear in the solar chromosphere.  
	
	2. In all the previous simulations, the EB was directly located in the reconnection region and heated by magnetic reconnection process. Though the small twisted flux rope with the EB originally comes into being by magnetic reconnection, it is a long distance from EB to reconnection region and the EB is heated by the shocks in this work as described in Section 3.1. Hence, we propose a new formation mechanism of EBs. 
	
	3. The formation heights of the EB and UV burst in this work indicate that both EBs and UV burst can appear in the low chromosphere, but the UV burst can extend to the upper chromosphere. The plasma $\beta$ in the EB region is obviously much larger than the UV burst, which are consistent with the previous 2D simulations \citep[e.g.,][]{Ni2021,Ni2016, Ni2018a,Peter2019}. 

The simplified radiative cooling term applied in our simulation matches well with the detailed losses in the VALC model. The initial distributions of plasma parameters in this work are similar to those in the C7 model (similar as the VALC model). Therefore, such a radiative cooling model is suitable at beginning of our simulation. However, the plasma parameters change a lot in the whole domain during the emergence process, especially in the reconnection region. Though the radiative cooling effect in our model varies with the temperature and plasma density, we do not know how far such a simple radiative cooling model caused the radiative cooling process to deviate from the realistic one, which needs further more realistic RMHD simulations to verify. The more realistic and natural radiation model in the future study must be also very requisite to cause the formation of a realistic convection zoom below the solar surface, which then naturally leads to more flux emerging processes and avoid the formation of the strip-like shape of the simulated EB. The losing of the chromospheric plasmas during the flux emerging process might also affect the synthesized H$\alpha$ images and spectral line profiles. Therefore, a higher simulation box in the further simulations is also very necessary. The non-equilibrium ionization effect is not included and the ionization degree does not vary with time in this work. The previous two-fluid simulations have shown that including the non-equilibrium ionization will make the temperature increase more difficult. However, the maximum temperature in the reconnection region can still reach above 20, 000 K if the reconnection magnetic fields around the solar TMR are stronger than 500 G. Therefore, the temperature increase in the EB and the UV burst might be overestimated in this work. If the temperature increase in the case by including the non-equilibrium ionization is lower, the corresponding contribution function for calculating the Si IV emissions in section 3.2 will be lower, but including the non-equilibrium ionization makes the number density of the electrons to be higher at the same time, it is hard to tell if the calculated Si IV emissions will be weaker or stronger than those presented in this work. On the other hand, we should point out that the more realistic magnetic diffusion might increase the heating and temperature in the chromosphere reconnection process, and the ambipolar diffusion \citep[e.g.,][]{Wang1993,Brandenburg1994,Ni2015} above the solar TMR might also supply more extra heating. The previous 2D high resolution simulations show that the turbulent multi-thermal structures and shocks appear inside the magnetic islands and reconnection outflow region \citep{Ni2021}, which are not obvious in our low resolution 3D simulation. Therefore, we also need further high resolution simulations to check these turbulent structures, the effects of a more realistic magnetic diffusion and ambipolar diffusion on the magnetic reconnection process in the future work. The stronger initial magnetic fields applied in the future simulations is also necessary to check if the broader Si IV spectral line profiles in UV bursts can be generated.

\normalem
\begin{acknowledgements}

The authors would like to thank the editor and the anonymous referee for the valuable comments to improve this work. This research is supported by NSFC Grants 11973083; the Strategic Priority Research Program of CAS with grants XDA17040507 and QYZDJ-SSWSLH012; the NSFC grant  11933009, and grants associated with the Yunling Scholar Project of the Yunnan Province and the Yunnan Province Scientist Workshop of Solar Physics; the Youth Innovation Promotion Association CAS 2017; the Applied Basic Research of Yunnan Province in China Grant 2018FB009; the Yunnan Ten-Thousand Talents Plan-Young top talents; the project of the Group for Innovation of Yunnan Province grant 2018HC023; the YunnanTen-Thousand Talents Plan-Yunling Scholar Project; the Special Program for Applied Research on Super Computation of the NSFC-Guangdong Joint Fund (nsfc2015-460, nsfc2015-463, the second phase); Computational Solar Physics Laboratory of Yunnan Observatories;the key Laboratory of Solar Activity grant KLSA202103.

\end{acknowledgements}
  
\bibliographystyle{raa}
\bibliography{bibtex}

\end{document}